\documentclass[aps, prd,
preprintnumbers,
longbibliography, 
nofootinbib, reprint
]{revtex4-2}

\usepackage[nowatermark]{fixmetodonotes}

\usepackage[T2A,T1]{fontenc}
\usepackage[utf8]{inputenc}
\usepackage[russian,american]{babel}

\usepackage{tipa}

\usepackage{graphicx}
\usepackage{isotope}
\usepackage{xcolor}

\makeatletter
\protected\def\xvcenter{%
  \hbox\bgroup$\everyvbox{\everyvbox{}\aftergroup\m@th\aftergroup$\aftergroup\egroup}%
  \vcenter
}

\DeclareRobustCommand{\midscript}[1]{
  \mathchoice{\mid@script\scriptstyle{#1}}
    {\mid@script\scriptstyle{#1}}
    {\mid@script\scriptscriptstyle{#1}}
    {\mid@script\scriptscriptstyle{#1}}
}
\newcommand{\mid@script}[2]{
  \vcenter{\hbox{$\m@th#1#2$}}
}

\DeclareRobustCommand{\textmidscript}[1]{%
  \xvcenter{\hbox{\scriptsize#1}}%
}
\makeatother

\newcommand{\cpp}{C\textmidscript{++}} 
\newcommand{\recoBinIdx}{a}
\newcommand{\trueBinIdx}{\mu}
\newcommand{\secondTrueBinIdx}{\lambda}
\newcommand{\secondRecoBinIdx}{b}
\newcommand{\PredictedRecoEvtCount}{n}
\newcommand{\PredictedSignalEvtCount}{\phi}
\newcommand{\MeasuredSignalEvtCount}{\hat{\PredictedSignalEvtCount}}

\newcommand{\PredictedBkgdCount}{B}
\newcommand{\ResponseMatrix}{\Delta}
\newcommand{\CovMat}{V}
\newcommand{\UnivIdx}{u}
\newcommand{\UnivCount}{N_{\mathrm{univ}}}
\newcommand{\MigrationMatrix}{M}
\newcommand{\UnfoldingMatrix}{U}
\newcommand{\ErrPropMatrix}{\mathfrak{E}}
\newcommand{\BkgdSubtractedRecoEvents}{d}
\newcommand{\AllRecoEvents}{D}
\newcommand{\AddSmearMatrix}{A_C}
\newcommand{\IterationIdx}{i}
\newcommand{\finalIterationIdx}{f}
\newcommand{\BinProb}{P}
\newcommand{\Eff}{\epsilon}

\newcommand{\XsecDimension}{n}
\newcommand{\SecondXsecDimension}{m}
\newcommand{\PhaseSpaceVec}{\mathbf{x}}
\newcommand{\SecondPhaseSpaceVec}{\mathbf{y}}
\newcommand{\PhaseSpaceVecWidths}{\Delta\PhaseSpaceVec}
\newcommand{\SecondPhaseSpaceVecWidths}{\Delta\SecondPhaseSpaceVec}
\newcommand{\IntegratedFlux}{\ensuremath{\Phi}}
\newcommand{\Flux}{\ensuremath{\varphi}}
\newcommand{\NumTargets}{T}
\newcommand{\DiffXsecAbbrev}{\ensuremath{s}}
\newcommand{\MeanXsecAbbrev}{\bar{\DiffXsecAbbrev}}
\newcommand{\Likelihood}{\mathcal{L}}
\newcommand{\LstatTerm}{\mathcal{P}}
\newcommand{\BBbeta}{\beta}
\newcommand{\BBsigma}{\sigma}
\newcommand{\systParamVec}{\mathbf{p}}
\newcommand{\systParamVecPrior}{\systParamVec^\mathrm{CV}}
\newcommand{\systParamVecPF}{\systParamVec^\mathrm{PF}}
\newcommand{\systParamCovMat}{V_\mathrm{syst}}
\newcommand{\sigScaleParam}{c}
\newcommand{\sigScaleParamPF}{\sigScaleParam^\mathrm{PF}}

\newcommand{\regStrength}{\tau_\mathrm{reg}}
\newcommand{\LCurveX}{\mathcal{X}}
\newcommand{\LCurveY}{\mathcal{Y}}
\newcommand{\HessianMatrix}{H}
\newcommand{\fitParam}{K}
\newcommand{\fitParamVec}{\mathbf{\fitParam}}
\newcommand{\fitParamVecPF}{\fitParamVec^{\mathrm{PF}}}
\newcommand{\fitParamIdx}{\text{\foreignlanguage{russian}{ч}}}
\newcommand{\secondFitParamIdx}{\text{\foreignlanguage{russian}{ъ}}}
\newcommand{\fluxParamVec}{\pmb{\psi}}
\newcommand{\nonFluxParamVec}{\pmb{\theta}}
\newcommand{\blockIdx}{\mathfrak{b}}
\newcommand{\bkgdScaleFactor}{\alpha}
\newcommand{\bkgdIdx}{\text{\textthorn}}

\newcommand{\randVec}{\mathbf{X}}
\newcommand{\randVecVal}{\mathbf{d}}
\newcommand{\meanVec}{\pmb{\mu}}
\newcommand{\randVecCovMat}{V}
\newcommand{\randVecFirstIdx}{1}
\newcommand{\randVecSecondIdx}{2}
\newcommand{\randVecCondLabel}{\text{constr}}

\newcommand{\signalRegionLabel}{S}
\newcommand{\controlSampleLabel}{C}

\newcommand{\auxCovMat}{V}

\newcommand{\sigVec}{\pmb{\PredictedSignalEvtCount}}
\newcommand{\bkgdVec}{\mathbf{\PredictedBkgdCount}}
\newcommand{\recoVec}{\mathbf{\PredictedRecoEvtCount}}
\newcommand{\recoWithUncVec}{\widetilde{\recoVec}}
\newcommand{\dataRecoVec}{\mathbf{\AllRecoEvents}}
\newcommand{\recoBkgdConstVec}{\pmb{\mathsf{m}}}
\newcommand{\recoBkgdConstWithUncVec}{\widetilde{\recoBkgdConstVec}}

\newcommand{\numRecoBins}{\ensuremath{R}}
\newcommand{\BBvec}{\ensuremath{\mathbf{F}}}
\newcommand{\BBsubVec}{\ensuremath{\mathbf{f}}}
\newcommand{\WSVDDiagMatrix}{\mathcal{D}_C}

\makeatletter
\def\@bibdataout@aps{%
\immediate\write\@bibdataout{%
@CONTROL{%
apsrev41Control%
\longbibliography@sw{%
    ,author="08",editor="1",pages="1",title="0",year="1"%
    }{%
    ,author="08",editor="1",pages="1",title="",year="1"%
    }%
  }%
}%
\if@filesw \immediate \write \@auxout {\string \citation {apsrev41Control}}\fi
}
\makeatother

\usepackage{setspace}

\usepackage{amsfonts}
\usepackage{amsmath}
\usepackage{amssymb}

\DeclareFontFamily{OT1}{pzc}{}
\DeclareFontShape{OT1}{pzc}{m}{it}{<-> s * [1.10] pzcmi7t}{}
\DeclareMathAlphabet{\mathpzc}{OT1}{pzc}{m}{it}

\usepackage{booktabs}
\AtBeginDocument{
\heavyrulewidth=.08em
\lightrulewidth=.05em
\cmidrulewidth=.03em
\belowrulesep=.65ex
\belowbottomsep=0pt
\aboverulesep=.4ex
\abovetopsep=0pt
\cmidrulesep=\doublerulesep
\cmidrulekern=.5em
\defaultaddspace=.5em
}

\usepackage{caption}
\usepackage{comment} 
\usepackage[inline]{enumitem}
\usepackage{graphicx}
\usepackage{isotope}
\usepackage{multirow} 

\usepackage{siunitx}
\DeclareSIUnit\ton{t}
\DeclareSIUnit\parsec{pc}
\DeclareSIUnit[number-unit-product = ]\percent{\char`\%}

\usepackage[colorlinks, urlcolor=blue, citecolor=., menucolor=.,
  anchorcolor=., linkcolor=., runcolor=.]{hyperref}

\begin{document}

\preprint{FERMILAB-PUB-23-692-CSAID}

\title{Mathematical methods for neutrino cross-section extraction}

\author{Steven Gardiner}
\email{gardiner@fnal.gov}
\affiliation{Fermi National Accelerator Laboratory, Batavia,
Illinois 60510 USA}

\date{\today}

\begin{abstract}
Precise modeling of neutrino-nucleus scattering is becoming increasingly
important as accelerator-based oscillation experiments seek definitive answers
to open questions about neutrino properties. To guide the needed model
refinements, a growing number of experimental collaborations are pursuing a
wide-ranging program of neutrino interaction measurements at \si{\GeV}
energies. A key step in most such analyses is cross-section extraction, in
which measured event counts are corrected for background contamination and
imperfect detector performance to yield cross-section results that are directly
comparable to theoretical predictions. In this paper, I review the major
approaches to cross-section extraction in the literature using representative
examples from the MINERvA, MicroBooNE, and T2K experiments. I then present two
mathematical techniques, blockwise unfolding and the conditional covariance
background constraint, which overcome some limitations of typical cross-section
extraction procedures.
\end{abstract}


\maketitle

\section{Introduction}
\label{sec:intro}

At the \si{\GeV} energies relevant for accelerator-based neutrino
oscillation experiments, there is increasing theoretical and experimental
investment toward achieving a precise understanding of the physics of neutrino
interactions with atomic nuclei~\cite{nf06report}. This effort is intended to
maximize the discovery potential of large, next-generation experiments like
Hyper-Kamiokande~\cite{HyperKdesign} and DUNE~\cite{DUNEtdrVol1}, which will
need percent-level control of systematic uncertainties to successfully execute
their flagship analyses~\cite{nustec}.

A rapidly growing literature of neutrino cross-section measurements is serving
as an indispensable resource for benchmarking calculations and improving
nuclear theory and simulations to the needed level. Due to the broad range of
energies produced in accelerator neutrino beams and the difficulty of
accurately reconstructing the incident neutrino energy on an event-by-event
basis, the majority of modern neutrino scattering measurements are presented as
flux-averaged differential cross sections.\footnote{As noted in a recent
review~\cite{Mahn2018}, the term \textit{flux-integrated} is also commonly used
in the literature, typically with an equivalent meaning.} These may generally
be written in the form
\begin{equation}
\label{eq:diff_xsec_flux_avg}
\left< \frac{ d^\XsecDimension\sigma }{ d\PhaseSpaceVec } \right>
 \equiv \frac{1}{\Phi} \int \Flux(E_\nu)
\, \frac{ d^\XsecDimension\sigma(E_\nu) }{ d\PhaseSpaceVec } \, dE_\nu \,,
\end{equation}
where
\begin{equation}
\Phi \equiv \int \Flux(E_\nu) \, dE_\nu
\end{equation}
is the integral of the beam flux $\Flux(E_\nu)$ over neutrino energy $E_\nu$,
and $d^\XsecDimension\sigma(E_\nu)/d\PhaseSpaceVec$ is the energy-dependent
differential cross section as a function of $\XsecDimension$ kinematic
variable(s) $\PhaseSpaceVec$ of interest.


The usual starting point for a neutrino cross-section analysis involves
defining the signal event topology and the observable(s) $\PhaseSpaceVec$ to be
measured. Based upon these choices, event selection criteria are developed to
isolate signal events from background. A binning scheme is also specified in
which selected events belonging to distinct ranges of the observables are
tallied separately from each other. Due to limited detector resolution, some
events will inevitably be assigned to an incorrect bin. The impact of these bin
migrations is estimated using simulation and typically quantified using two
separate sets of bins. The \textit{true bins} are defined in terms of the
actual values of the observables, while the \textit{reconstructed bins} collect
events whose measured values fall within an appropriate range.

The final step of a neutrino scattering analysis is known as
\textit{cross-section extraction} and involves correcting the observed event
counts in each reconstructed bin for remaining background, detector efficiency,
and bin migrations. Scaling factors are then applied to the corrected event
counts to obtain cross sections that can be compared to interaction model
predictions.

While important and sometimes subtle differences exist in the cross-section
extraction procedures used by different experimental collaborations, nearly all
strategies currently in use can be expressed in the form
\begin{equation}
\label{eq:extract}
\left< \frac{ d^\XsecDimension\sigma }{ d\PhaseSpaceVec }
\right>_{\!\trueBinIdx} = \frac{ \sum_\recoBinIdx
\UnfoldingMatrix_{\trueBinIdx \recoBinIdx}
\, ( \AllRecoEvents_\recoBinIdx - \PredictedBkgdCount_\recoBinIdx ) }
{ \IntegratedFlux \, \NumTargets \, \PhaseSpaceVecWidths_\trueBinIdx } \,.
\end{equation}
Here the symbol $\AllRecoEvents_\recoBinIdx$
($\PredictedBkgdCount_\recoBinIdx$) represents the total number of measured
events (estimated number of background events) in the $\recoBinIdx$-th
reconstructed bin.

The elements $\UnfoldingMatrix_{\trueBinIdx \recoBinIdx}$ of the
\textit{unfolding matrix} are used to apply efficiency and bin migration
corrections to the background-subtracted event counts. They may equivalently be
written as
\begin{equation}
\label{eq:unfolding_matrix}
\UnfoldingMatrix_{\trueBinIdx \recoBinIdx}
= \frac{ \BinProb_{\trueBinIdx \recoBinIdx} }{ \Eff_\trueBinIdx } \,,
\end{equation}
where $\Eff_\trueBinIdx$ is the detection efficiency for signal events
belonging to the $\trueBinIdx$-th true bin, and $\BinProb_{\trueBinIdx
\recoBinIdx}$ is the conditional probability that a signal event measured
within the $\recoBinIdx$-th reconstructed bin also belongs to the
$\trueBinIdx$-th true bin. In this paper, Greek subscripts ($\trueBinIdx$,
$\secondTrueBinIdx$) are used to represent true bin indices, while Latin
subscripts ($\recoBinIdx$, $\secondRecoBinIdx$) represent reconstructed bin
indices.

The denominator from Eq.~\ref{eq:extract} contains the scaling factors needed
to convert the unfolded signal event count into a differential cross section.
The symbol $T$ denotes the number of scattering targets illuminated by the
neutrino beam and included in the active region of the detector used to measure
the reconstructed event counts $\AllRecoEvents_\recoBinIdx$. Expressing $T$ as
a number of target atoms or a number of target nucleons are both common
conventions in the literature. The symbol $\PhaseSpaceVecWidths_\trueBinIdx$
represents the product of the $\XsecDimension$ bin widths for the
$\trueBinIdx$-th true bin.

The resulting differential cross section on the left-hand side of
Eq.~\ref{eq:extract} is an average value in the $\trueBinIdx$-th true bin. If
this bin is defined so that it contains all signal events with values of the
observables $\PhaseSpaceVec$ such that
\begin{equation}
\label{eq:true_bin_def}
\PhaseSpaceVec \in [\PhaseSpaceVec_\trueBinIdx,
\PhaseSpaceVec_{\trueBinIdx + 1}) \,,
\end{equation}
then the average cross section in this bin may be written in terms of the
definition given in Eq.~\ref{eq:diff_xsec_flux_avg} as
\begin{equation}
\label{eq:diff_xsec_flux_avg_in_bin}
\left< \frac{ d^\XsecDimension\sigma }{ d\PhaseSpaceVec }
\right>_{\!\trueBinIdx} \equiv \frac{1}{\PhaseSpaceVecWidths_\trueBinIdx}
\int_{\PhaseSpaceVec_\trueBinIdx}^{\PhaseSpaceVec_{\trueBinIdx + 1}}
\left< \frac{ d^\XsecDimension\sigma }{ d\PhaseSpaceVec } \right>
\, d\PhaseSpaceVec \,.
\end{equation}
The $\XsecDimension$-dimensional limits $\PhaseSpaceVec_\trueBinIdx$ and
$\PhaseSpaceVec_{\trueBinIdx + 1}$ define the region of kinematic phase space
spanned by the bin.

\section{Current approaches}
\label{sec:current_approaches}

Significant variations exist between the approaches to cross-section extraction
employed in modern neutrino experiments. While the basic concepts described in
the introduction are essentially universal, key differences exist between the
methods used to estimate unfolding corrections and to compute uncertainties on
the extracted measurement. To provide sufficient context for the mathematical
techniques presented in later sections of this paper, three representative
cross-section extraction strategies from the MINERvA, MicroBooNE, and T2K
experiments are described below.

\subsection{MINERvA-style extraction}
\label{sec:minerva_extract}

The MINERvA collaboration has published an extensive library of flux-averaged
neutrino cross sections~\cite{MINERvA:2023ner, Olivier2023, Kleykamp2023,
MINERvA:2023avz, MINERvA:2022esg, MINERvA:2022djk, MINERvA:2022mnw,
MINERvA:2022bno, MINERvA:2021wjs, MINERvA:2021owq, MINERvA:2020zzv,
MINERvA:2020anu, MINERvA:2019gsf, MINERvA:2019ope, MINERvA:2019rhx,
MINERvA:2018hqn, MINERvA:2018hba, MINERvA:2018vjb, MINERvA:2017ipy,
MINERvA:2017okh, MINERvA:2017dzh, MINERvA:2016ymg, MINERvA:2016sfc,
MINERvA:2016zyp, MINERvA:2016oql, MINERvA:2015ydy, MINERvA:2015jih,
MINERvA:2015slz, MINERvA:2014ypj, MINERvA:2014ani, MINERvA:2014ogb,
MINERvA:2013bcy, MINERvA:2013kdn} following a consistent extraction procedure.
Unfolding corrections are evaluated using Richardson-Lucy
deconvolution~\cite{Richardson:72,Lucy:1974yx}, an iterative algorithm
popularized in high-energy physics by D'Agostini~\cite{Dagostini1995}. A
MINERvA-specific fork of the RooUnfold~\cite{Adye:2011gm} software package,
called UnfoldUtils~\cite{unfoldutils}, is coupled to the MINERvA Analysis
Toolkit~\cite{MINERvA:2021ddh} to provide the numerical implementation.

\subsubsection{D'Agostini unfolding}
\label{sec:dagostini_unfold}
The starting point for the D'Agostini method is a set of initial estimators
$\MeasuredSignalEvtCount_\trueBinIdx^{0}$ for the unfolded signal event counts
(i.e., the numerator in Eq.~\ref{eq:extract}) in each of the $\trueBinIdx$ true
bins. These are arbitrary but typically taken from the main simulation
prediction used to execute the analysis. Each iteration $i$ of the method
obtains an updated estimator
\begin{equation}
\label{eq:updated_estimator_dagostini}
\MeasuredSignalEvtCount_\trueBinIdx^{\IterationIdx + 1}
= \sum_\recoBinIdx \UnfoldingMatrix_{\trueBinIdx \recoBinIdx}^{\IterationIdx}
\, \BkgdSubtractedRecoEvents_\recoBinIdx
\end{equation}
by applying the unfolding matrix from Eq.~\ref{eq:unfolding_matrix} to the
background-subtracted measured event counts
\begin{equation}
\label{eq:bkgd_subtract}
\BkgdSubtractedRecoEvents_\recoBinIdx \equiv D_\recoBinIdx - B_\recoBinIdx
\end{equation}
in each of the $\recoBinIdx$ reconstructed bins. While the detection efficiency
$\Eff_\trueBinIdx$ is held constant over iterations, the conditional
probability is updated each time via the formula
\begin{equation}
\label{eq:binprob_dagostini}
\BinProb^{\IterationIdx}_{\trueBinIdx \recoBinIdx}
= \frac{ \MigrationMatrix_{\recoBinIdx \trueBinIdx}
\, \MeasuredSignalEvtCount_\trueBinIdx^{\IterationIdx} }
{ \sum_\secondTrueBinIdx \MigrationMatrix_{\recoBinIdx \secondTrueBinIdx}
\, \MeasuredSignalEvtCount_\secondTrueBinIdx^{\IterationIdx} } \,.
\end{equation}
Here the elements of the \textit{migration matrix}
$\MigrationMatrix_{\recoBinIdx \trueBinIdx}$ represent the probability that a
measured signal event belonging to the $\trueBinIdx$-th true bin will be
assigned to the $\recoBinIdx$-th reconstructed bin. They are typically
estimated from simulation via
\begin{equation}
\label{eq:migmat}
\MigrationMatrix_{\recoBinIdx \trueBinIdx} =
\frac{ \PredictedSignalEvtCount_{\recoBinIdx\trueBinIdx} }
{ \sum_\secondRecoBinIdx
\PredictedSignalEvtCount_{\secondRecoBinIdx\trueBinIdx} }
\end{equation}
where $\PredictedSignalEvtCount_{\recoBinIdx\trueBinIdx}$ is the number of
simulated signal events belonging simultaneously to the $\trueBinIdx$-th true
bin and the $\recoBinIdx$-th reconstructed bin.

The total number of iterations $\finalIterationIdx$ is typically chosen based
on the unfolding performance in mock-data studies. The estimator
$\MeasuredSignalEvtCount_\trueBinIdx^{\finalIterationIdx}$ obtained in the last
iteration is used to obtain the final cross-section result.

\subsubsection{Multiple-universe extraction}
\label{sec:multi_universe_extract}
Systematic uncertainties on the measurement are represented as covariance
matrices that are calculated using a set of $\UnivCount$ alternative
simulations, or \textit{universes}. In each universe $\UnivIdx$, the quantities
$\UnfoldingMatrix_{\trueBinIdx \recoBinIdx}$,
$\PredictedBkgdCount_\recoBinIdx$, $\IntegratedFlux$, and $\NumTargets$ are
calculated based upon the alternative simulation, and the cross section is
re-extracted according to Eq.~\ref{eq:extract}. The covariance on the resulting
differential cross section is then computed as the maximum likelihood estimate
assuming that the universes are drawn from a multivariate Gaussian
distribution:
\begin{equation}
\label{eq:minerva_multisim}
\mathrm{Cov}( \DiffXsecAbbrev_\trueBinIdx,
\DiffXsecAbbrev_\secondTrueBinIdx ) = \frac{1}{\UnivCount}
\sum^{\UnivCount}_{u=1} (\DiffXsecAbbrev_\trueBinIdx^\UnivIdx
- \MeanXsecAbbrev_\trueBinIdx)
(\DiffXsecAbbrev_\secondTrueBinIdx^\UnivIdx
- \MeanXsecAbbrev_\secondTrueBinIdx) \,.
\end{equation}
Here I have defined the abbreviation
\begin{equation}
\label{eq:abbrev_diff_xsec}
\DiffXsecAbbrev_\trueBinIdx \equiv
\left< \frac{ d^\XsecDimension\sigma }{ d\PhaseSpaceVec }
\right>_{\!\trueBinIdx} \,.
\end{equation}
The symbol $\DiffXsecAbbrev_\trueBinIdx^\UnivIdx$ denotes the value of the
extracted differential cross section $\DiffXsecAbbrev_\trueBinIdx$ evaluated in
the $\UnivIdx$-th universe. For a single alternative universe ($\UnivCount =
1$), the symbol $\MeanXsecAbbrev_\trueBinIdx$ represents the value of
$\DiffXsecAbbrev_\trueBinIdx$ obtained using the nominal MINERvA simulation.
For multiple systematic variations, $\MeanXsecAbbrev_\trueBinIdx$ denotes the
arithmetic mean of $\DiffXsecAbbrev_\trueBinIdx$ evaluated in all of the
universes.

Detailed documentation for the treatment of data statistical uncertainties in
the MINERvA procedure does not appear to be available in their publications.
However, a bootstrapping method involving re-extraction of the cross sections
$\DiffXsecAbbrev_\trueBinIdx$ in universes for which the measured event counts
$\AllRecoEvents_\recoBinIdx$ fluctuate according to Poisson statistics has been
used in a MINERvA-style analysis by NOvA~\cite{nova2dnumu}. Analytic
propagation of the statistical covariance matrix through unfolding, as
described for MicroBooNE in Sec.~\ref{sec:analytic_prop}, would also be a
reasonable approach.

\subsubsection{Real-flux measurement}
\label{sec:real_flux_minerva}
As discussed at length in Ref.~\cite{KochDolan}, the extraction procedure
adopted by MINERvA leads to a differential cross section result that is
averaged over the \textit{real} neutrino flux received by the detector. This
flux differs from the \textit{reference} flux obtained from simulation and used
to compute theoretical predictions of the cross sections in a way that is not
exactly known. Rigorously quantifying goodness-of-fit (e.g., via a chi-squared
metric) between a cross-section calculation and the data extracted according to
the MINERvA treatment therefore requires uncertainties on the flux shape as a
function of neutrino energy to be evaluated on the model prediction.

\subsubsection{Use by other experiments}
MINERvA-like strategies for cross-section extraction are dominant in the
accelerator neutrino interaction literature, with similar approaches being used
in measurements by NOvA~\cite{NOvA:2024rov, NOvA:2024zmr, nova2dnumu,
nova2dnue, novapi0}, T2K~\cite{T2K:2020txr, T2K:2019yqu, T2K:2019dgm,
T2K:2018rnz, T2K:2017qxv, T2K:2016cbz, T2K:2016jor, T2K:2014lbi,T2K:2013nor},
and MiniBooNE~\cite{MiniBooNE:2009dxl, MiniBooNE:2010xqw, MiniBooNE:2010bsu,
MiniBooNE:2010cxl, MiniBooNE:2010eis, MiniBooNE:2013qnd, MiniBooNE:2013dds}.
Several cross-section analyses from ArgoNeuT and one from MicroBooNE evaluate
uncertainties in the MINERvA style by re-extracting the cross section in
multiple systematic universes, but bin migration corrections are either
entirely neglected~\cite{ArgoNeuT:2011bms, ArgoNeuT:2014rlj} or are evaluated
in a more approximate way using effective values of the detection
efficiency~\cite{ArgoNeuT:2020kir, Fitzpatrick:2021egx, ArgoNeuT:2018und,
Abratenko:2020acr}.

\subsection{MicroBooNE-style extraction}
\label{sec:ub_style}

In an effort to mitigate concerns about model dependence in the extraction
procedure, two early MicroBooNE cross-section measurements were reported using
a \textit{forward-folding} approach~\cite{mcc8CCincl, mcc8ccnppaper}. Under
this strategy, bin migration adjustments are applied to theoretical predictions
rather than the extracted data when performing quantitative comparisons.
Limitations of the MicroBooNE forward-folding formalism were examined in
subsequent community discussions~\cite{Avanzini:2021qlx} that discouraged
further use. Although tools exist that would enable a more sophisticated
variation of forward-folding~\cite{Koch2019}, the MicroBooNE collaboration has
opted to release unfolded results in all recent publications presenting
differential cross sections. Most of these~\cite{nueCCDiffXSec,
WireCellCCinclPRL, tkiprl, tkiprd, microboone:2023foc, uBGKI,
MicroBooNE:2024zwf, MicroBooNE:2024zkh, MicroBooNE:2024sec, MicroBooNE:2024yxe}
relatively new unfolding technique called the Wiener-SVD
method~\cite{WienerSVD}, which was originally implemented in a dedicated
code~\cite{wienersvdrepo} based on ROOT~\cite{brun1997, rootwebsite}.

\subsubsection{Wiener-SVD unfolding}
\label{sec:wsvd}

The Wiener-SVD technique is built on the observation that unfolding the signal
event counts may be interpreted as a likelihood optimization problem. A
straightforward approach uses direct inversion of the \textit{detector response
matrix}
\begin{equation}
\label{eq:response_matrix}
\ResponseMatrix_{\recoBinIdx \trueBinIdx}
\equiv \Eff_\trueBinIdx \, \MigrationMatrix_{\recoBinIdx \trueBinIdx}
\end{equation}
to obtain the unfolding matrix:
\begin{equation}
\label{eq:unfolding_direct}
\UnfoldingMatrix^\mathrm{direct} = ( \ResponseMatrix^T \ResponseMatrix )^{-1}
\, \ResponseMatrix^T \,.
\end{equation}
Here the left inverse is written rather than the ordinary inverse
($\Delta^{-1}$) to cover cases in which $\ResponseMatrix$ is not a square
matrix. While $\UnfoldingMatrix^\mathrm{direct}$ corresponds to the maximum
likelihood solution to the unfolding problem, its use has known pathologies,
including large uncertainties. To improve upon the direct inversion strategy,
standard unfolding techniques rely upon \textit{regularization}: new
constraints are introduced into the likelihood function based on prior
information about the expected solution. These are intended to reduce the
variance on the unfolded result at the cost of some (hopefully small) bias.

The Wiener-SVD authors note that the impact of regularization can be written
for a general unfolding matrix $\UnfoldingMatrix$ in terms of the direct
inversion one via
\begin{equation}
\label{eq:add_smear_in_unfolding}
\UnfoldingMatrix = \AddSmearMatrix \cdot \UnfoldingMatrix^\mathrm{direct} \,.
\end{equation}
Thus, an unfolding method that can be represented as a matrix transformation of
the background-subtracted data, as in Eq.~\ref{eq:extract}, may be expressed as
a recipe for construction of the \textit{regularization matrix} or
\textit{additional smearing matrix} $\AddSmearMatrix$. A full derivation of
$A_C$ for the Wiener-SVD method, which is based on an analogy with the Wiener
filter used in signal processing, is provided in the original
publication~\cite{WienerSVD}. The details of the calculation are unimportant
for the present discussion, but Appendix~\ref{sec:Ac_wsvd} provides a brief
summary.

\subsubsection{Analytic uncertainty propagation}
\label{sec:analytic_prop}

One of the required inputs for the Wiener-SVD unfolding method is a covariance
matrix $\mathrm{Cov}( \AllRecoEvents_\recoBinIdx,
\AllRecoEvents_\secondRecoBinIdx )$ that includes all statistical and
systematic uncertainties on the measured total event counts
$\AllRecoEvents_\recoBinIdx$ in the reconstructed bins $\recoBinIdx$. The need
for this covariance matrix motivates an uncertainty treatment for MicroBooNE
that is distinct from MINERvA's. The differences in the uncertainty
quantification strategy are ultimately more consequential than those in the
specifics of the unfolding itself.

Under the MicroBooNE approach, data statistical covariances are evaluated in
the usual way for Poisson statistics. Systematic covariances are estimated
using the corresponding covariances on the expected event counts
$\PredictedRecoEvtCount_\recoBinIdx$ obtained from simulation:
\begin{equation}
\mathrm{Cov}( \AllRecoEvents_\recoBinIdx, \AllRecoEvents_\secondRecoBinIdx )
\approx \mathrm{Cov}( \PredictedRecoEvtCount_\recoBinIdx,
\PredictedRecoEvtCount_\secondRecoBinIdx ) \,.
\end{equation}
These in turn are computed using a multiple-universe approach similar to the
one in Eq.~\ref{eq:minerva_multisim} from MINERvA:
\begin{equation}
\label{eq:CovMat}
\mathrm{Cov}(\PredictedRecoEvtCount_\recoBinIdx,
\PredictedRecoEvtCount_\secondRecoBinIdx) =
\frac{ 1 }{ \UnivCount } \sum_{\UnivIdx = 1}^{\UnivCount} \big(
\PredictedRecoEvtCount_{\recoBinIdx}^{\UnivIdx}
- \PredictedRecoEvtCount_{\recoBinIdx}^\mathrm{CV} \big) \big(
\PredictedRecoEvtCount_{\secondRecoBinIdx}^{\UnivIdx} -
\PredictedRecoEvtCount_{\secondRecoBinIdx}^\mathrm{CV} \big) \,.
\end{equation}
In this case, however, the quantity of interest is the simulation prediction
for the number of events $\PredictedRecoEvtCount_\recoBinIdx$ in the
$\recoBinIdx$-th reconstructed bin, and the \textit{central-value} prediction
from MicroBooNE's nominal simulation
$\PredictedRecoEvtCount_{\recoBinIdx}^\mathrm{CV}$ is used in the expression
rather than the mean value from the alternate universes.

A special treatment is employed for uncertainties arising from MicroBooNE's
neutrino interaction model~\cite{microboonegenietune}. While the expected event
counts $\PredictedRecoEvtCount_{\recoBinIdx}$ are are varied directly in most
of the systematic universes, those related to the interaction model use the
expression
\begin{equation}
\label{eq:ExpectedRecoEvents}
\PredictedRecoEvtCount_{\recoBinIdx} = \PredictedBkgdCount_{\recoBinIdx}
+ \sum_{\trueBinIdx} \ResponseMatrix_{\recoBinIdx \trueBinIdx} \,
\PredictedSignalEvtCount_{\trueBinIdx}^\mathrm{CV}
\end{equation}
in which the detector response matrix $\ResponseMatrix_{\recoBinIdx
\trueBinIdx}$ is applied to the central-value prediction of signal event counts
in each of the $\trueBinIdx$ true bins. Because uncertainties related to
interaction modeling of signal events will affect cross-section extraction only
through their influence on unfolding, the background
$\PredictedBkgdCount_\recoBinIdx$ and the response matrix elements
$\ResponseMatrix_{\recoBinIdx \trueBinIdx}$ are varied in each relevant
universe while the signal prediction
$\PredictedSignalEvtCount_{\trueBinIdx}^\mathrm{CV}$ is held constant. When the
central-value response matrix is used, the second term in
Eq.~\ref{eq:ExpectedRecoEvents} is just the nominal prediction for the number
of signal events in the $\recoBinIdx$-th reconstructed bin.

Since background subtraction involves shifting the content of each
reconstructed bin by a constant, the covariance on the background-subtracted
event counts $\BkgdSubtractedRecoEvents_\recoBinIdx$ from
Eq.~\ref{eq:bkgd_subtract} is the same as the covariance before subtraction:
\begin{equation}
\mathrm{Cov}( \BkgdSubtractedRecoEvents_\recoBinIdx,
\BkgdSubtractedRecoEvents_\secondRecoBinIdx )
= \mathrm{Cov}( \AllRecoEvents_\recoBinIdx,
\AllRecoEvents_\secondRecoBinIdx ) \,.
\end{equation}

Final uncertainties on the differential cross section measurement are obtained
by analytically propagating the covariance matrices through the extraction
procedure, which is applied only once. Using the abbreviated notation from
Eq.~\ref{eq:abbrev_diff_xsec}, the covariance matrix describing the measurement
may be written as
\begin{equation}
\label{eq:ub_xsec_unc}
\mathrm{Cov}( \DiffXsecAbbrev_\trueBinIdx, \DiffXsecAbbrev_\secondTrueBinIdx )
= \frac{ \mathrm{Cov}(\MeasuredSignalEvtCount_\trueBinIdx,
\MeasuredSignalEvtCount_\secondTrueBinIdx) }
{ \IntegratedFlux^2 \, \NumTargets^2 \, \PhaseSpaceVecWidths_\trueBinIdx
\, \PhaseSpaceVecWidths_\secondTrueBinIdx } \,.
\end{equation}
Here the covariance on the unfolded signal event counts
\begin{equation}
\label{eq:unfold}
\MeasuredSignalEvtCount_\trueBinIdx = \sum_\recoBinIdx
\UnfoldingMatrix_{\trueBinIdx \recoBinIdx}
\, \BkgdSubtractedRecoEvents_\recoBinIdx
\end{equation}
is computed via
\begin{equation}
\label{eq:unfold_prop}
\mathrm{Cov}(\MeasuredSignalEvtCount_\trueBinIdx,
\MeasuredSignalEvtCount_\secondTrueBinIdx)
= \sum_{\recoBinIdx, \secondRecoBinIdx}
\ErrPropMatrix_{\trueBinIdx \recoBinIdx}
\,
\mathrm{Cov}( \BkgdSubtractedRecoEvents_\recoBinIdx,
\BkgdSubtractedRecoEvents_\secondRecoBinIdx )
\,
\ErrPropMatrix^T_{\secondRecoBinIdx \secondTrueBinIdx} \,.
\end{equation}
In general, the elements of the \textit{error propagation matrix}
$\ErrPropMatrix$ are the partial derivatives
\begin{equation}
\label{eq:err_prop_matrix_general}
\ErrPropMatrix_{\trueBinIdx \recoBinIdx} \equiv
\frac{ \partial \MeasuredSignalEvtCount_\trueBinIdx }
{ \partial \BkgdSubtractedRecoEvents_\recoBinIdx } \,.
\end{equation}
For Wiener-SVD and similar methods where the unfolding matrix does not depend
on the data, this becomes simply
\begin{equation}
\label{eq:err_prop_equals_unfolding}
\ErrPropMatrix_{\trueBinIdx \recoBinIdx} =
\UnfoldingMatrix_{\trueBinIdx \recoBinIdx} \,.
\end{equation}

\subsubsection{Reference-flux measurement}
\label{sec:ref_flux_uboone}
In contrast to MINERvA-style cross-section extraction, the MicroBooNE approach
produces a measurement in the \textit{reference} flux estimated from the
nominal simulation. This is a consequence of the choice to directly vary the
expected event counts $\PredictedRecoEvtCount_{\recoBinIdx}$ in the flux
systematic universes rather than using the prescription from
Eq.~\ref{eq:ExpectedRecoEvents} in which signal events are only adjusted
through variations in the detector response matrix
$\ResponseMatrix_{\recoBinIdx \trueBinIdx}$. A key assumption in this choice is
that the neutrino energy dependence of the total cross section $\sigma(E_\nu)$
from the nominal simulation is sufficiently realistic to be used to propagate
beam flux uncertainties into the expected event counts
$\PredictedRecoEvtCount_{\recoBinIdx}$.

This also leads to a subtle change in the interpretation of the measured event
counts $\AllRecoEvents_\recoBinIdx$. These are observed numbers of events
measured in the real flux, and they are interpreted as such under a
MINERvA-like approach. However, the MicroBooNE uncertainty prescription treats
them as estimators of the event counts that \textit{would have} been measured
in the reference flux. Since the reference flux used is the best available
estimate of the real flux, no correction is needed to the values of the
$\AllRecoEvents_\recoBinIdx$ in the analysis.

Because the full effect of the flux and target-counting systematic variations
is already included in the pre-unfolding covariance matrix, the denominator in
Eq.~\ref{eq:ub_xsec_unc} uses the central-value estimates of $\IntegratedFlux$
and $\NumTargets$ without assigning any additional uncertainty.

\subsubsection{Use of the regularization matrix}

To account for the bias introduced by unfolding methods that use
regularization, occasionally an ad hoc systematic uncertainty is added to an
analysis. For example, because the D'Agostini method approaches direct
inversion in the limit of many iterations, an implicit regularization is
applied by the choice of the total iteration count $\finalIterationIdx$. Some
neutrino cross-section analyses have therefore taken the spread between the
results obtained with different numbers of iterations as an additional
uncertainty on the unfolding procedure itself~\cite{MiniBooNE:2010bsu,
MiniBooNE:2013dds}.

Although well-intended, application of this extra uncertainty decreases the
statistical power of the measurement unnecessarily. The authors of
Ref.~\cite{WienerSVD} recommend instead that the regularization matrix
$\AddSmearMatrix$ be reported together with the unfolded cross-section results.
If one multiplies a theoretical prediction in the true bins by
$\AddSmearMatrix$ before a goodness-of-fit metric is calculated, then the
regularization is applied to the prediction and the data in a consistent way,
avoiding the need for any additional uncertainty. The degree to which
$\AddSmearMatrix$ differs from the identity matrix also provides a convenient
way of quantifying the level of bias introduced by the chosen regularization.

In situations where Eq.~\ref{eq:err_prop_equals_unfolding} holds, use of
$\AddSmearMatrix$ makes the $\chi^2$ goodness-of-fit score invariant with
respect to the unfolding procedure. That is, if one forward-folds a vector of
predicted event counts $\pmb{\PredictedSignalEvtCount}$ via multiplication by
the response matrix $\ResponseMatrix$ and compares the result to the
background-subtracted data $\mathbf{\BkgdSubtractedRecoEvents}$ in the
reconstructed bins, one obtains exactly the same $\chi^2$ statistic as when one
multiplies $\pmb{\PredictedSignalEvtCount}$ by $\AddSmearMatrix$ and compares
to $\pmb{\MeasuredSignalEvtCount}$, the unfolded measurement:
\begin{align}
\label{eq:chi_squared_AC_1}
\chi^2 &= ( \mathbf{\BkgdSubtractedRecoEvents}
- \ResponseMatrix \cdot \pmb{\PredictedSignalEvtCount} )^{T}
\cdot (\CovMat^\mathrm{reco})^{-1} \cdot
( \mathbf{\BkgdSubtractedRecoEvents}
- \ResponseMatrix \cdot \pmb{\PredictedSignalEvtCount} ) \\
\label{eq:chi_squared_AC_2}
&= ( \pmb{\MeasuredSignalEvtCount}
- \AddSmearMatrix \cdot \pmb{\PredictedSignalEvtCount} )^{T}
\cdot (\CovMat^\mathrm{unf})^{-1} \cdot
( \pmb{\MeasuredSignalEvtCount}
- \AddSmearMatrix \cdot \pmb{\PredictedSignalEvtCount} ) \,.
\end{align}
Here the reconstructed ($\CovMat^\mathrm{reco}$) and unfolded
($\CovMat^\mathrm{unf}$) covariance matrices respectively have elements
\begin{align}
\CovMat^\mathrm{reco}_{\recoBinIdx \secondRecoBinIdx}
&\equiv \mathrm{Cov}(\AllRecoEvents_\recoBinIdx,
\AllRecoEvents_\secondRecoBinIdx) \\
\CovMat^\mathrm{unf}_{\trueBinIdx \secondTrueBinIdx}
&\equiv \mathrm{Cov}( \MeasuredSignalEvtCount_\trueBinIdx,
\MeasuredSignalEvtCount_\secondTrueBinIdx) \,,
\end{align}
and are related via Eq.~\ref{eq:unfold_prop}, which may be written in matrix
form as
\begin{equation}
\label{eq:unfold_prop_matrix_form}
\CovMat^\mathrm{unf}
= \ErrPropMatrix \cdot \CovMat^\mathrm{reco} \cdot \ErrPropMatrix^T \,.
\end{equation}
Thus, a presentation strategy in which goodness-of-fit with the unfolded data
is calculated after multiplying model predictions by $\AddSmearMatrix$ can
retain the statistical power of forward-folding while providing more intuitive
unfolded results for display in plots and qualitative comparisons to
theory~\cite{Koch:2022qsz}.

All MicroBooNE cross-section publications to date that have used Wiener-SVD
unfolding have followed this data reporting strategy~\cite{nueCCDiffXSec,
WireCellCCinclPRL, tkiprl, tkiprd, microboone:2023foc, uBGKI,
MicroBooNE:2024zwf, MicroBooNE:2024zkh, MicroBooNE:2024sec,
MicroBooNE:2024yxe}. However, the calculation of $\AddSmearMatrix$ is simple
for any unfolding method expressible as a linear transformation of the data.
For an arbitrary unfolding matrix $\UnfoldingMatrix$, it follows immediately
from the definitions in Eqs.~\ref{eq:unfolding_direct} and
\ref{eq:add_smear_in_unfolding} that
\begin{equation}
\label{eq:add_smear_general}
\AddSmearMatrix = \UnfoldingMatrix \cdot \ResponseMatrix \,.
\end{equation}
For MINERvA-like uncertainty treatments in which the unfolding matrix is
recalculated in each systematic universe, the central-value version should be
used to obtain $\AddSmearMatrix$ for use in a data release.

\subsubsection{Uncertainty propagation for D'Agostini}

When using the D'Agostini method of unfolding (Sec.~\ref{sec:dagostini_unfold})
together with analytic uncertainty propagation, as has been done in two
MicroBooNE cross-section analyses so far~\cite{BNBnueXsec, MicroBooNE:2024tmp},
the matrix transformation in Eqs.~\ref{eq:unfold_prop}
and~\ref{eq:unfold_prop_matrix_form} becomes more subtle. Applying the
definition of the error propagation matrix elements
(Eq.~\ref{eq:err_prop_matrix_general}) to the expression for the unfolded
signal event counts from Eq.~\ref{eq:unfold} yields
\begin{equation}
\label{eq:err_prop_chain_rule}
\ErrPropMatrix_{\trueBinIdx \recoBinIdx} =
\UnfoldingMatrix_{\trueBinIdx \recoBinIdx} +
\sum_{\secondRecoBinIdx}
\BkgdSubtractedRecoEvents_\secondRecoBinIdx \,
\frac{ \partial \UnfoldingMatrix_{\trueBinIdx \secondRecoBinIdx} }
{ \partial \BkgdSubtractedRecoEvents_\recoBinIdx }
\end{equation}
via the chain rule. Because elements of the unfolding matrix $\UnfoldingMatrix$
depend on the measured event counts $\mathbf{\BkgdSubtractedRecoEvents}$ when
multiple iterations $\IterationIdx$ are used, the second term in
Eq.~\ref{eq:err_prop_chain_rule} no longer trivially vanishes as it does for
the Wiener-SVD method. A correct treatment of this term depends on whether the
regularization matrix $\AddSmearMatrix$ will be used in the presentation of the
unfolded result.

In the traditional case where $\AddSmearMatrix$ is not used in model
comparisons to the data, the unfolding matrix $\UnfoldingMatrix$ represents an
estimate of the transformation needed to move from the reconstructed event
distribution to the true one. This transformation is imperfectly known, and the
sum in Eq.~\ref{eq:err_prop_chain_rule} represents the extent to which the
specific estimate chosen $\UnfoldingMatrix$ depends on the reconstructed
measurement $\mathbf{\BkgdSubtractedRecoEvents}$ obtained in the analysis. For
multiple D'Agostini iterations, a different measurement
$\mathbf{\BkgdSubtractedRecoEvents}$ would yield a different unfolding matrix
$\UnfoldingMatrix$, and this source of unfolding-related uncertainty must
therefore be included via the sum over $\secondRecoBinIdx$. In Sec.~5.1 of
Ref.~\cite{Adye:2011gm}, the need to account for this unfolding uncertainty is
directly demonstrated using a toy MC study.

An explicit form for the error propagation matrix elements needed to propagate
uncertainties for D'Agostini unfolding without the use of $\AddSmearMatrix$ is
derived in Refs.~\cite{Bourbeau2019,Bourbeau2018} and may be written as
\begin{equation}
\label{eq:errprop_dagostini}
\ErrPropMatrix_{\trueBinIdx \recoBinIdx}^{\IterationIdx + 1} =
\frac{ \partial \MeasuredSignalEvtCount_\trueBinIdx^{\IterationIdx+1} }
{ \partial \BkgdSubtractedRecoEvents_\recoBinIdx } =
\UnfoldingMatrix_{\trueBinIdx \recoBinIdx}^{\IterationIdx}
+ \frac{ \MeasuredSignalEvtCount_\trueBinIdx^{\IterationIdx + 1} }
{ \MeasuredSignalEvtCount^i_\trueBinIdx }
\ErrPropMatrix_{\trueBinIdx \recoBinIdx}^{\IterationIdx}
- \sum_{\secondTrueBinIdx,\secondRecoBinIdx}
\Eff_\secondTrueBinIdx \, \frac{ \BkgdSubtractedRecoEvents_\secondRecoBinIdx }
{ \MeasuredSignalEvtCount^\IterationIdx_\secondTrueBinIdx }
\, \UnfoldingMatrix_{\trueBinIdx \secondRecoBinIdx}^{\IterationIdx}
\, \UnfoldingMatrix_{\secondTrueBinIdx \secondRecoBinIdx}^{\IterationIdx} \,
\ErrPropMatrix_{\secondTrueBinIdx \recoBinIdx}^{\IterationIdx}
\end{equation}
with $\ErrPropMatrix_{\trueBinIdx \recoBinIdx}^{0} = 0$. The covariance matrix
describing the uncertainty on the unfolded event counts obtained from the final
iteration $\finalIterationIdx$ is given by
\begin{equation}
\label{eq:unfold_prop_dagostini}
\mathrm{Cov}(\MeasuredSignalEvtCount_\trueBinIdx^\finalIterationIdx,
\MeasuredSignalEvtCount_\secondTrueBinIdx^\finalIterationIdx)
= \sum_{\recoBinIdx, \secondRecoBinIdx}
\ErrPropMatrix_{\trueBinIdx \recoBinIdx}^\finalIterationIdx
\,
\mathrm{Cov}( \BkgdSubtractedRecoEvents_\recoBinIdx,
\BkgdSubtractedRecoEvents_\secondRecoBinIdx )
\,
\ErrPropMatrix_{\secondTrueBinIdx \secondRecoBinIdx}^\finalIterationIdx \,.
\end{equation}
This treatment of uncertainty propagation for D'Agostini unfolding was used by
MicroBooNE in Ref.~\cite{BNBnueXsec}, which reports differential cross sections
without providing a regularization matrix.

When $\AddSmearMatrix$ is used for quantitative model comparisons to data
unfolded via the D'Agostini method, as was done in
Ref.~\cite{MicroBooNE:2024tmp} by MicroBooNE, a different interpretation of the
error propagation matrix $\ErrPropMatrix$ becomes relevant, and the expression
in Eq.~\ref{eq:err_prop_chain_rule} must be revisited. As discussed in
Ref.~\cite{Koch:2022qsz}, the unfolding procedure may now be viewed as a
convenient aid for data visualization that preserves all information from the
original measurement $\mathbf{\BkgdSubtractedRecoEvents}$.

Although the unfolding matrix $\UnfoldingMatrix$ is still constructed based on
an (imperfect) estimate of the transformation required to obtain true event
counts from the reconstructed ones, this is now done solely because the results
thus become easier to interpret qualitatively as an approximation to the true
distribution. While a different measurement
$\mathbf{\BkgdSubtractedRecoEvents}$ could have lead to a different unfolding
matrix $\UnfoldingMatrix$, only the matrix actually used in the analysis is
relevant. This is because a consistent linear transformation is applied to
model predictions (via multiplication by $\AddSmearMatrix$) for quantitative
comparisons. The second term in Eq.~\ref{eq:err_prop_chain_rule} therefore
becomes superfluous and should be neglected; there is no new uncertainty
introduced by the unfolding operation itself. A helpful consequence of this
approach is that Eq.~\ref{eq:err_prop_equals_unfolding} becomes satisfied for
D'Agostini unfolding as it is for the Wiener-SVD method, and one obtains the
same $\chi^2$ value from forward-folding (Eq.~\ref{eq:chi_squared_AC_1}) as
from unfolding (Eq.~\ref{eq:chi_squared_AC_2}).

\subsection{T2K-style extraction}
\label{sec:t2k_extraction}

Like MicroBooNE, the T2K collaboration has developed a unique cross-section
extraction procedure that has become the preferred method in their recent
analyses~\cite{T2KCohPiPlus, T2KCC0piTwoDetector, T2K:2021naz, T2K:2020jav,
T2K:2020lrr, T2K:2020sbd, T2K:2019ddy, T2K:2018lnf}. In two early
publications~\cite{T2K:2018rnz,T2K:2016jor}, the new T2K technique was used to
report some results while others were obtained using a MINERvA-style approach.

The core of the T2K strategy is a binned likelihood fit to the data involving a
large number of parameters. This fit is used for unfolding, background removal,
and at least the first stage of uncertainty quantification. The Minuit2
code~\cite{minuit2}, a \cpp\ translation of the Fortran software package
MINUIT~\cite{James:1975dr}, provides the algorithms used for numerical
optimization. An initial implementation of the T2K fitting framework is
publicly available in the form of the xsLLhFitter~\cite{xsllhFitter} code.
However, further software development effort appears to be focused instead on a
forked version called GUNDAM~\cite{gundam}.

\subsubsection{Binned likelihood fit}

For the T2K fitting framework, the task of interest is minimization
of minus two times the logarithm of the likelihood function:
\begin{equation}
\label{eq:log_likelihood_t2k}
-2\log(\Likelihood) = -2\log(\Likelihood_\mathrm{stat})
- 2\log(\Likelihood_\mathrm{syst}) - 2\log(\Likelihood_\mathrm{reg}) \,.
\end{equation}
This quantity is identical to the chi-squared statistic ($\chi^2$) when all
parameters of interest follow a multivariate Gaussian distribution. This is
true for the $\Likelihood_\mathrm{syst}$ term but not the others.

\paragraph{Statistical term} The statistical contribution to the log-likelihood
may be written as
\begin{equation}
-2\log(\Likelihood_\mathrm{stat}) = 2\sum_\recoBinIdx
\LstatTerm_\recoBinIdx \,,
\end{equation}
where the sum runs over all reconstructed bins $\recoBinIdx$.
In the earliest versions of the T2K fitter, $\LstatTerm_\recoBinIdx$
took the simple form~\cite{T2K:2016jor}
\begin{equation}
\label{eq:poisson_likelihood}
\LstatTerm_\recoBinIdx = \PredictedRecoEvtCount_\recoBinIdx
- \AllRecoEvents_\recoBinIdx + \AllRecoEvents_\recoBinIdx
\log\!\bigg(\frac{ \AllRecoEvents_\recoBinIdx }
{ \PredictedRecoEvtCount_\recoBinIdx }\bigg)
\end{equation}
where $\AllRecoEvents_\recoBinIdx$ ($\PredictedRecoEvtCount_\recoBinIdx$) is
the measured (predicted) number of events in the $\recoBinIdx$-th reconstructed
bin. This expression is the negative log-likelihood for a Poisson distribution
after applying Stirling's approximation~\cite{stirling}
\begin{equation}
\log(\AllRecoEvents_\recoBinIdx!)
\approx \AllRecoEvents_\recoBinIdx \log(\AllRecoEvents_\recoBinIdx)
- \AllRecoEvents_\recoBinIdx \,.
\end{equation}
More recent updates to the T2K fitter described in Refs.~\cite{T2K:2021naz,
T2KCC0piTwoDetector} have added corrections to the \mbox{Poissonian} likelihood
for finite Monte Carlo (MC) statistics in the calculation of the
$\PredictedRecoEvtCount_\recoBinIdx$ using the Barlow-Beeston
approach~\cite{Prosper:2011zz,Barlow1993}. This yields a new expression
\begin{equation}
\LstatTerm_\recoBinIdx = \BBbeta_\recoBinIdx \PredictedRecoEvtCount_\recoBinIdx
- \AllRecoEvents_\recoBinIdx + \AllRecoEvents_\recoBinIdx
\log\!\bigg(\frac{ \AllRecoEvents_\recoBinIdx }{ \BBbeta_\recoBinIdx
\PredictedRecoEvtCount_\recoBinIdx }\bigg) + \frac{ \BBbeta^2_\recoBinIdx - 1 }
{ 2\BBsigma^2_\recoBinIdx }
\end{equation}
where $\BBsigma^2_\recoBinIdx$ is the relative MC statistical variance of
$\PredictedRecoEvtCount_\recoBinIdx$ and
\begin{equation}
\BBbeta_\recoBinIdx = \frac{1}{2}
\left[ 1 - \PredictedRecoEvtCount_\recoBinIdx \BBsigma_\recoBinIdx^2
+ \sqrt{ (\PredictedRecoEvtCount_\recoBinIdx \BBsigma_\recoBinIdx^2 - 1 )^2
+ 4\AllRecoEvents_\recoBinIdx \BBsigma_\recoBinIdx^2 }
\right] \,.
\end{equation}
In the limit of infinite simulated events, the expression in
Eq.~\ref{eq:poisson_likelihood} is recovered via $\BBsigma_\recoBinIdx^2 \to
0$.

\paragraph{Systematic term} The systematic contribution in
Eq.~\ref{eq:log_likelihood_t2k} takes the form
\begin{equation}
\label{eq:log_likelihood_syst}
-2\log(\Likelihood_\mathrm{syst}) =
(\systParamVec - \systParamVecPrior)^{T} \cdot \systParamCovMat^{-1} \cdot
(\systParamVec - \systParamVecPrior)
\end{equation}
and acts as a Gaussian penalty term that prevents the systematic parameters
$\systParamVec$ varied in the fit from deviating implausibly far from their
fixed central values $\systParamVecPrior$. The covariance matrix
$\systParamCovMat$ describes the prior uncertainties on the systematic
parameters, including their correlations.

The predicted events $\PredictedRecoEvtCount_\recoBinIdx$ in each reconstructed
bin are calculated as a function of the systematic parameters via
\begin{equation}
\PredictedRecoEvtCount_\recoBinIdx
= \sum_\trueBinIdx \big[ \sigScaleParam_\trueBinIdx \,
\PredictedSignalEvtCount_{\trueBinIdx \recoBinIdx}(\systParamVec)
+ \PredictedBkgdCount_{\trueBinIdx \recoBinIdx}(\systParamVec)
\big]
\end{equation}
where $\PredictedSignalEvtCount_{\trueBinIdx \recoBinIdx}$
($\PredictedBkgdCount_{\trueBinIdx \recoBinIdx}$) is the simulated number of
signal (background) events that fall simultaneously into the $\trueBinIdx$-th
true bin and the $\recoBinIdx$-th reconstructed bin. An additional parameter
$\sigScaleParam_\trueBinIdx$ is included in the fit for each true bin
$\trueBinIdx$. Since the $\sigScaleParam_\trueBinIdx$ are unconstrained by the
systematic log-likelihood in Eq.~\ref{eq:log_likelihood_syst}, the signal
prediction is allowed to float in the fit to match the data.

Due to the large computational cost of running the simulation repeatedly, the
dependence of the signal ($\PredictedSignalEvtCount_{\trueBinIdx \recoBinIdx}$)
and background ($\PredictedBkgdCount_{\trueBinIdx \recoBinIdx}$) predictions on
the systematic parameters $\systParamVec$ is evaluated via reweighting. Under
this approach, a large set of MC events is first generated using the
nominal simulation. To obtain a prediction for an alternative simulation
($\systParamVec \neq \systParamVecPrior$), a statistical weight is assigned to
each event based on its relative likelihood of occurring in the systematic
universe of interest.

\paragraph{Regularization term} Minimization of the log-likelihood formed by
the first two terms from Eq.~\ref{eq:log_likelihood_t2k} is similar to directly
inverting the detector response matrix (see Sec.~\ref{sec:wsvd}) and gives an
equivalent result under certain conditions~\cite[Sec.\,IV\,D\,1]{T2K:2018rnz}.
While the best-fit values of the parameters $\sigScaleParam_\trueBinIdx$ and
$\systParamVec$ are obtained with minimal bias, the procedure suffers from the
same difficulties as direct inversion. In particular, the fit tends to be
highly sensitive to small statistical fluctuations in the measured event counts
$\AllRecoEvents_\recoBinIdx$ and to yield a corresponding estimator of the
observed signal event counts with strong negative correlations between
neighboring bins.

Similarly to the D'Agostini and Wiener-SVD unfolding strategies, which address
these concerns by modifying direct matrix inversion using prior information
about the expected result, regularization can also be applied in the T2K
fitting technique. In this case, the prior information is represented by an
optional third term in the log-likelihood function. The form of this third term
is ultimately arbitrary and may be defined in an analysis-specific way.
However, a choice used in multiple T2K publications~\cite{T2K:2020jav,
T2K:2019ddy, T2K:2018rnz} is a version of Tikhonov-Phillips
regularization~\cite{TikhonovReg,PhillipsReg} in which a smoothness condition
is imposed on the signal scaling parameters $\sigScaleParam_\trueBinIdx$ via
\begin{equation}
-2\log(\Likelihood_\mathrm{reg}) = \regStrength
\sum_{\trueBinIdx} (\sigScaleParam_{\trueBinIdx + 1}
- \sigScaleParam_{\trueBinIdx})^2 \,.
\end{equation}
Here the sum runs over pairs of neighboring true bins $\trueBinIdx$ and
\mbox{$\trueBinIdx + 1$} representing the same observable. The quantity in
parentheses is proportional to the first derivative of the
$\sigScaleParam_\trueBinIdx$ under a forward difference approximation.

The regularization strength $\regStrength$ is a constant that controls the
relative importance of this term in the likelihood fit. In the T2K analyses
adopting regularization of this kind, the value of $\regStrength$ was chosen
using using an \textit{L-curve} technique~\cite{HansenLCurve}. This approach
seeks an optimal regularization strength by repeating the fit many times for
different $\regStrength$ values. Each fit result contributes a point to a
parametric curve in which the $x$ and $y$ coordinates are given by
\begin{equation}
\LCurveX = -2\log(\Likelihood_\mathrm{stat})
  - 2\log(\Likelihood_\mathrm{syst}) \,,
\end{equation}
and
\begin{equation}
\LCurveY =
\frac{-2 \log(\Likelihood_\mathrm{reg}) }{ \regStrength } \,.
\end{equation}
Both coordinates are evaluated after minimizing the full log-likelihood
$-2\log(\Likelihood)$ at fixed $\regStrength$. Because $\LCurveX$ tends to
increase and $\LCurveY$ tends to decrease as the regularization strength grows,
the parametric curve $(\LCurveX, \LCurveY)$ is expected to have a
characteristic \textit{L} shape. Choosing the $\regStrength$ value
corresponding to the point of maximum curvature (i.e., the kink in the
\textit{L}) thus represents an optimal balance between compatibility with
the regularization criterion (minimal $\LCurveY$) and compatibility with the
measured data (minimal $\LCurveX$).

\paragraph{Fit results} To minimize the log-likelihood function defined in
Eq.~\ref{eq:log_likelihood_t2k}, the T2K fitter uses the MIGRAD
algorithm~\cite{T2KCohPiPlus,James:1975dr} as implemented in Minuit2. The
unfolded signal event counts $\MeasuredSignalEvtCount_\trueBinIdx$ in each true
bin $\trueBinIdx$ are then calculated after minimization via
\begin{equation}
\label{eq:t2k_unfolded_signal}
\MeasuredSignalEvtCount_\trueBinIdx =
\frac{ 1 }{ \Eff_\trueBinIdx }
\sum_\recoBinIdx \sigScaleParamPF_\trueBinIdx \,
\PredictedSignalEvtCount_{\trueBinIdx \recoBinIdx}(\systParamVecPF) \,,
\end{equation}
where the sum includes all relevant reconstructed bins $\recoBinIdx$.

This leads to the measured differential cross section
\begin{equation}
\label{eq:t2k_diff_xsec}
\DiffXsecAbbrev_\trueBinIdx = \frac{ \MeasuredSignalEvtCount_\trueBinIdx }
{ \IntegratedFlux \, \NumTargets \, \PhaseSpaceVecWidths_\trueBinIdx } \,.
\end{equation}
where the abbreviation defined in Eq.~\ref{eq:abbrev_diff_xsec} is used once
again. The superscript $^\mathrm{PF}$ that appears in
Eq.~\ref{eq:t2k_unfolded_signal} denotes the post-fit values of the parameters.
The efficiency $\Eff_\trueBinIdx$, integrated flux $\IntegratedFlux$, and
number of scattering targets $\NumTargets$ are also evaluated using the
post-fit parameters. For example, the efficiency can be estimated from the T2K
simulation as
\begin{equation}
\Eff_\trueBinIdx = \frac{ \sum_\recoBinIdx
\PredictedSignalEvtCount_{\trueBinIdx \recoBinIdx}(\systParamVecPF) }
{ \PredictedSignalEvtCount_{\trueBinIdx}(\systParamVecPF) }
\end{equation}
where $\PredictedSignalEvtCount_{\trueBinIdx}$ is the total number of simulated
signal events in the $\trueBinIdx$-th true bin.

\subsubsection{Fit parameter covariances}

In addition to providing a means of extracting the measured differential cross
sections $\DiffXsecAbbrev_\trueBinIdx$, the log-likelihood function from
Eq.~\ref{eq:log_likelihood_t2k} encodes detailed information about the
relationships between all parameters included in the fit. Using the HESSE
algorithm~\cite{T2KCohPiPlus, James:1975dr} from Minuit2, an approximate
covariance matrix describing uncertainties on the fit parameters may be
obtained by calculating the elements of the \textit{Hessian matrix}
\begin{equation}
\label{eq:t2k_hessian}
\HessianMatrix_{\fitParamIdx \secondFitParamIdx}
= \left. - \frac{ \partial^2 \log(\Likelihood) }
{ \partial \fitParam_\fitParamIdx \, \partial \fitParam_\secondFitParamIdx}
\right|_{ \fitParamVec = \fitParamVecPF }
\end{equation}
and then inverting it:
\begin{equation}
\label{eq:t2k_param_covariance}
\mathrm{Cov}( \fitParam_\fitParamIdx, \fitParam_\secondFitParamIdx )
\approx (\HessianMatrix^{-1})_{ \fitParamIdx \secondFitParamIdx } \,.
\end{equation}
Here the signal scaling factors $\sigScaleParam_\trueBinIdx$ and the systematic
parameters included in $\systParamVec$ all appear as individual elements
$\fitParam_\fitParamIdx$ of a combined vector of fit parameters
\begin{equation}
\fitParamVec = \begin{pmatrix}
c_0 \\
c_1 \\
\vdots \\
\systParamVec
\end{pmatrix} \,.
\end{equation}
Cyrillic subscripts ($\fitParamIdx$, $\secondFitParamIdx$) are used as
parameter indices to avoid confusion with those used to identify the true and
reconstructed bins. The post-fit values $\fitParamVecPF$ define the point in
parameter space where the Hessian matrix elements $\HessianMatrix_{\fitParamIdx
\secondFitParamIdx}$ are evaluated. The approximate equality in
Eq.~\ref{eq:t2k_param_covariance} becomes exact when all of the
$\fitParam_\fitParamIdx$ follow a multivariate Gaussian
distribution~\cite{HessianCovariance} and when the second partial derivatives
in Eq.~\ref{eq:t2k_hessian} are exactly calculable.

\subsubsection{Uncertainty propagation}
\label{sec:t2k_unc_prop}

Two distinct strategies have been used in T2K analyses for assigning an
uncertainty to the differential cross sections $\DiffXsecAbbrev_\trueBinIdx$
measured using a binned likelihood fit.

In Ref.~\cite[Sec.~IV\,E\,1]{T2K:2020jav}, a MINERvA-like multiple-universe
extraction approach (see Sec.~\ref{sec:multi_universe_extract} herein) is
adopted. In this case, a single set of universes is generated by varying all
systematic parameters simultaneously, and the fitting procedure is repeated in
each universe $\UnivIdx$ to yield a new measurement
$\DiffXsecAbbrev^{\UnivIdx}_\trueBinIdx$. The arithmetic mean of the results
$\MeanXsecAbbrev_\trueBinIdx$ is taken to be the final measured cross section,
and the covariance matrix describing its uncertainty is calculated according to
Eq.~\ref{eq:minerva_multisim}. Statistical uncertainties are included in the
multiple-universe treatment by fluctuating the contents of the reconstructed
bins according to a Poisson distribution.

In Refs.~\cite{T2KCC0piTwoDetector, T2K:2021naz}, repetition of the likelihood
fit is avoided by use of the parameter covariance matrix  mentioned above. Each
of the $\UnivCount$ universes is now generated by sampling a set of parameter
values $\fitParamVec^\UnivIdx$ from the multivariate Gaussian distribution with
mean $\fitParamVecPF$ and covariance matrix $\mathrm{Cov}(
\fitParam_\fitParamIdx, \fitParam_\secondFitParamIdx )$. The final measured
cross section is taken to be the value calculated with the post-fit parameters
$ \DiffXsecAbbrev_\trueBinIdx^\mathrm{PF}
\equiv \DiffXsecAbbrev_\trueBinIdx(\fitParamVecPF)$, and the elements of
the corresponding covariance matrix are obtained via
\begin{equation}
\label{eq:CovMatFit}
\mathrm{Cov}( \DiffXsecAbbrev_\trueBinIdx, \DiffXsecAbbrev_\secondTrueBinIdx )
= \frac{ 1 }{ \UnivCount } \sum_{\UnivIdx = 1}^{\UnivCount} \big(
\DiffXsecAbbrev_\trueBinIdx^\UnivIdx
- \DiffXsecAbbrev_\trueBinIdx^\mathrm{PF} \big) \big(
\DiffXsecAbbrev_\secondTrueBinIdx^\UnivIdx
- \DiffXsecAbbrev_\secondTrueBinIdx^\mathrm{PF}
\big) \,,
\end{equation}
where $\DiffXsecAbbrev_\trueBinIdx^\UnivIdx =
\DiffXsecAbbrev_\trueBinIdx(\fitParamVec^\UnivIdx)$ is the differential cross
section evaluated in the $\UnivIdx$-th universe. Note that the values of
$\Eff_\trueBinIdx$, $\IntegratedFlux$, and $\NumTargets$ are updated in each
universe to be consistent with the other parts of the calculation in
Eqs.~\ref{eq:t2k_unfolded_signal}~and~\ref{eq:t2k_diff_xsec}.

\subsubsection{Flux treatment}

The standard T2K method for cross-section extraction produces a real-flux
result with the same requirements for properly treating the flux shape
uncertainty as MINERvA's measurements (see Sec.~\ref{sec:real_flux_minerva}).
However, one recent T2K analysis~\cite{T2KCohPiPlus} reported a reference-flux
cross section instead by applying an extrapolation discussed in
Ref.~\cite{KochDolan}, which I summarize here.

Without loss of generality, arrange the elements of the vector of systematic
parameters $\systParamVec$ so that
\begin{equation}
\systParamVec =
\begin{pmatrix} \fluxParamVec \\[0.5mm] \nonFluxParamVec \end{pmatrix} \,,
\end{equation}
where the vector $\fluxParamVec$ contains the subset of parameters that affect
the neutrino flux model, while $\nonFluxParamVec$ contains those that do not.
To extract a cross-section measurement in a specific reference flux, the binned
likelihood fit and, if needed, the parameter covariance calculation from
Eq.~\ref{eq:t2k_param_covariance} are carried out unaltered. However, when
evaluating the differential cross section in a systematic universe
($\DiffXsecAbbrev_\trueBinIdx^\UnivIdx$) or at the post-fit point
($\DiffXsecAbbrev_\trueBinIdx^\mathrm{PF}$), the values of the flux parameters
$\fluxParamVec$ that would normally be used are ignored and replaced with
constants corresponding to the reference flux: $\fluxParamVec \to
\fluxParamVec^\mathrm{ref}$. This replacement applies to all component parts of
the cross-section calculation, including $\Eff_\trueBinIdx$ and
$\IntegratedFlux$. Thus, as is the case for the MicroBooNE extraction procedure
(see Sec.~\ref{sec:ref_flux_uboone}), the measurement becomes an estimator for
the cross section that \textit{would have been} observed in the reference flux.

\subsubsection{Regularization matrix}

Although a regularization matrix does not appear to have been provided yet as
part of any T2K cross-section data release, $\AddSmearMatrix$ has been shown to
be easily calculable using the unregularized likelihood fit results together
with Tikhonov-Phillips regularization~\cite{Koch:2022qsz}. A post-hoc
regularization may thus be applied to a T2K-style analysis without any need to
rerun the fit. This capability could allow a regularized cross-section result
obtained with a likelihood fit to be reported while preserving (via the use of
$\AddSmearMatrix$ in goodness-of-fit calculations) the statistical power of the
unregularized version.

\section{Blockwise unfolding}

In recent years, there has been growing recognition of the importance of
correlated uncertainties in the interpretation of neutrino cross-section data,
particularly for quantitative comparisons and model parameter tuning. While
early cross-section results from MiniBooNE established many analysis techniques
that remain influential for modern experiments, a notable but not
universal~\cite{MiniBooNE:2010xqw, MiniBooNE:2009dxl, MiniBooNE:2010cxl}
omission in some of the collaboration's data releases~\cite{miniboonedata} is
the lack of a full covariance matrix describing bin-to-bin correlations in the
measurement uncertainties~\cite{MiniBooNE:2013dds, MiniBooNE:2013qnd,
MiniBooNE:2010bsu, MiniBooNE:2010eis}. Such correlations arise from both
systematic and statistical effects, the latter as a result of unfolding even
when the reconstructed bins are disjoint. Difficulties created by the missing
MiniBooNE correlations in assessing agreement of predictions with the
data~\cite{Betancourt2018,Avanzini:2021qlx} and in fitting model
parameters~\cite{Wilkinson2016, TenaVidal2022} led to widespread consensus on
the need for this information and multiple calls for its inclusion in future
results~\cite{nustec, Katori2018}. Problems with missing correlations of this
kind are not limited to neutrino data sets; efforts to constrain uncertainties
on intranuclear cascade models using hadron-nucleus data have also had to
confront them~\cite{FSItune}.

Happily, data reporting strategies in the field were quick to adapt in light of
these limitations, and presenting a flux-averaged differential cross section
together with a corresponding uncertainty covariance matrix is now standard
practice. Although the value of this step forward should not be understated, a
primary goal of the present work is to recommend and enable the inclusion of
still more detailed information about measurement correlations in cross-section
data releases by neutrino experiments.

\subsection{Missing correlations}

Specifically, there are two types of covariances that are usually not reported
but nevertheless highly valuable for data interpretation. The first type arises
for measurements in which multiple kinematic distributions are presented,
typically using a consistent signal definition, event selection, and data set
in the analysis. A recent example is Refs.~\cite{tkiprd, tkiprl} from
MicroBooNE, in which various one- and two-dimensional differential cross
sections are measured using a single set of 9051 quasielastic-like event
candidates. Each differential cross section reported therein thus represents a
particular projection of a higher-dimensional joint distribution describing the
relationship between all relevant kinematic variables. In the absence of a
measurement of the full joint distribution, the most stringent test of an
interaction model is obtained by requiring it to describe all of the
low-dimensional projections simultaneously. However, inter-projection
covariances, i.e., those correlating a bin from one measured differential cross
section with a bin from another, are rarely provided in the literature. This
presents a major problem when attempting to quantify overall goodness-of-fit
between a model prediction and a multi-projection data set; strong
inter-distribution correlations are expected from both systematic uncertainties
(e.g., beam flux modeling) and statistical fluctuations (shared events).

A second type of covariance that is typically unavailable is evaluated between
cross-section results obtained by distinct analyses from the same experiment.
Many of the motivations and challenges for reporting this information are
shared with the first type, but there are additional practical concerns, such
as a possibly long time delay between execution of the two measurements.
Careful planning to ensure that inter-analysis covariances can be evaluated,
however, would do much to overcome difficulties encountered by multiple groups
examining neutrino scattering data over the past decade. In
Refs.~\cite{Wilkinson2016} and \cite{Bonus2020}, for example, presumably strong
correlations between neutrino and antineutrino cross sections measured
separately by the same experiment had to be neglected due to the lack of a
suitable data release.\footnote{Note, however, that the T2K data studied in
Ref.~\cite{Bonus2020} included such correlations while the MINERvA data did
not.}

Missing covariances of both types posed a challenge for a model fitting
study~\cite{MINERvAPionTune} in which the MINERvA collaboration tuned
parameters in the GENIE event generator~\cite{genie:2021npt,genie} to four of
their published pion production cross-section
measurements~\cite{MINERvA:2014ogb, MINERvA:2016sfc, MINERvA:2017okh}.
Inter-distribution correlations (first type) were handled with an ad hoc
approach to calculating the $\chi^2$ statistic, while inter-analysis
correlations (second type) were neglected entirely. The authors are forthright
in the paper about the limitations of these approximations, which were adopted
despite unfettered access to MINERvA's data and analysis tools. This suggests
that, even when an experiment has dedicated substantial effort to data
preservation~\cite{Fine2022}, retroactively obtaining correlations between
existing analyses may be prohibitively difficult in many cases.

\subsection{Precedents}

Since retrofitting existing results with more detailed uncertainty
quantification may not be feasible, one may instead design future neutrino
cross-section analyses in a way that facilitates reporting complete covariances
of both types. In the remainder of this section, I describe a mathematical
procedure intended to accomplish this goal.

Some elements of this procedure appear to have already been applied in recent
T2K analyses that made simultaneous measurements of neutrino and antineutrino
cross sections~\cite{T2K:2020sbd} and neutrino cross sections on both carbon
and oxygen targets~\cite{T2K:2020jav}. However, in that context, the primary
motivation was robust separation of signal and background, and the presentation
is entirely specific to a T2K-style cross-section extraction. Several MINERvA
publications have also reported simultaneous measurements of neutrino cross
sections on multiple targets~\cite{MINERvA:2017dzh, MINERvA:2022djk,
Kleykamp2023}. Although these publications report uncertainty covariance
matrices only for each individual distribution and its ratio to the
corresponding distribution for hydrocarbon, a proper handling of the
uncertainties for the latter suggests that a full set of covariances between
hydrocarbon and the other targets must have been obtained at some stage of the
analysis. Since the uncertainty treatment is not described in detail, however,
it is difficult to generalize to other applications.

With the aim of encouraging experimental collaborations to further strengthen
their neutrino cross-section data releases, I build upon these precedents to
develop a general recipe called \textit{blockwise unfolding} for evaluating
covariances between distinct flux-averaged differential cross section
measurements. I also discuss this method's possible application to evaluating
covariances between results obtained by separate analyses from the same
experiment. This includes identifying the specific information that must be
preserved to avoid time-consuming repetition of an earlier analysis for this
purpose.

The blockwise unfolding technique is compatible with at least some versions of
all three cross-section extraction strategies reviewed in
Sec.~\ref{sec:current_approaches}. Since incorporating blockwise unfolding
within a MicroBooNE-style measurement is the case that requires the most
detailed explanation, I assume it for an initial presentation below.
Adjustments appropriate for MINERvA- and T2K-style cross-section extraction are
then considered as an expansion of the prior material.

\subsection{MicroBooNE-style measurements}

As discussed in Sec.~\ref{sec:ub_style}, MicroBooNE's cross-section extraction
method involves evaluating a complete set of uncertainties on the number of
events measured in each reconstructed bin. A final result is obtained by
applying consistent linear transformations to both the background-subtracted
data and the total covariance matrix describing the reconstructed event counts.
To generalize the procedure to allow reporting the covariances between bins in
distinct kinematic distributions, two significant problems must be solved.
First, the covariance matrix describing the statistical uncertainty on the
measured event counts must be constructed in a way that respects the
correlations that arise due to events that are shared between distributions.
Second, the linear transformation applied to the data, which amounts to
unfolding and division by a few constant scaling factors, must be performed in
a way that preserves the inter-distribution covariances evaluated between
reconstructed bins.

\subsubsection{Statistical correlations}
\label{sec:stat_corr}

In the traditional case where neutrino cross sections are reported individually
for each kinematic distribution, calculation of the statistical covariance
matrix is simple. Each measured event belongs to exactly one reconstructed bin,
and hence the individual bin contents before unfolding follow independent
Poisson distributions. The statistical covariance matrix is diagonal, and the
number of observed counts $\AllRecoEvents_\recoBinIdx$ in the $\recoBinIdx$-th
reconstructed bin is an estimator of its variance.

\begin{figure}
\centering
\includegraphics[page=1, width=0.48\textwidth]{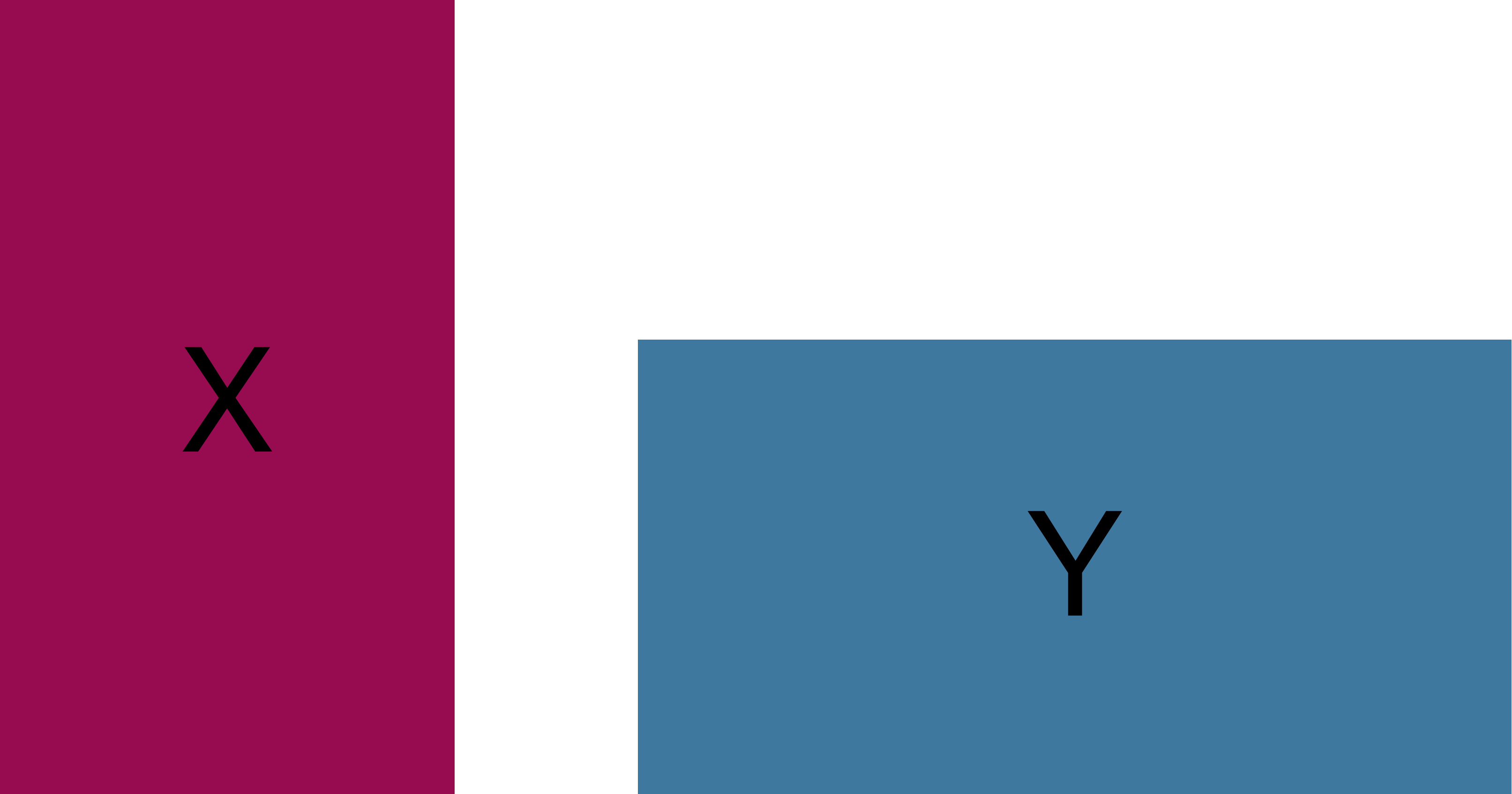}
\\[0.3cm]
\includegraphics[page=1, width=0.3\textwidth]{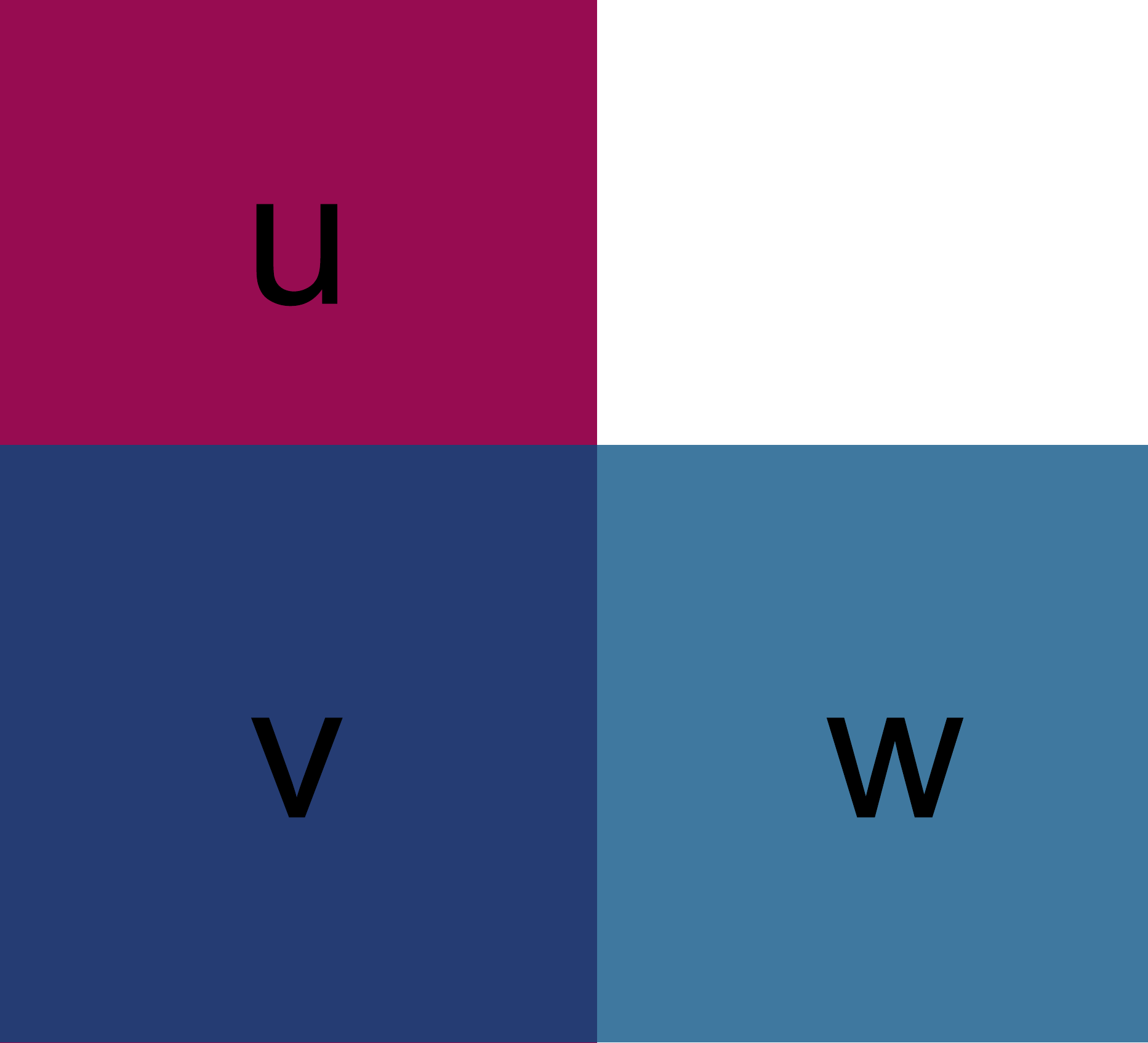}
\caption{A pair of arbitrary kinematic bins $X$ and $Y$ (top) can always be
subdivided into three non-overlapping bins $\mathrm{u}$, $\mathrm{v}$, and
$\mathrm{w}$ (bottom). This observation allows a straightforward derivation of
the statistical covariance between bins $X$ and $Y$.}
\label{fig:corr_stat_errors}
\end{figure}

When multiple distributions are extracted simultaneously from the same data
set, events can be shared between bins, and the assumption of independent
Poisson fluctuations in each bin is no longer valid. However, a general
expression for an estimator of the statistical covariance between an arbitrary
pair of kinematic bins, which I will label $X$ and $Y$, is easy to derive when
the problem is framed suitably.

As illustrated in Fig.~\ref{fig:corr_stat_errors}, the key trick is to subdivide
bins $X$ and $Y$ into three non-overlapping bins $\mathrm{u}$, $\mathrm{v}$, and
$\mathrm{w}$. Bin $\mathrm{v}$ is defined to contain only those events that
belong simultaneously to bins $X$ and $Y$, while bin $\mathrm{u}$ ($\mathrm{w}$)
contains the events that belong solely to bin $X$ ($Y$). By definition, then,
bins $\mathrm{u}$, $\mathrm{v}$, and $\mathrm{w}$ have no events in common, and
the measured event counts in each follow independent Poisson distributions. It
follows from the descriptions above that the covariance between the measured
event counts in bin $X$ and bin $Y$ is given by
\begin{align}
\nonumber
\mathrm{Cov}( \AllRecoEvents_X, \AllRecoEvents_Y )
&= \mathrm{Cov}( \AllRecoEvents_\mathrm{u} + \AllRecoEvents_\mathrm{v}, \AllRecoEvents_\mathrm{v} +
\AllRecoEvents_\mathrm{w} ) \\
\nonumber
&= \mathrm{Cov}( \AllRecoEvents_\mathrm{u}, \AllRecoEvents_\mathrm{v} )
+ \mathrm{Cov}( \AllRecoEvents_\mathrm{u}, \AllRecoEvents_\mathrm{w} )
\\ \nonumber & \;\;\;\;\;\;
+ \mathrm{Cov}( \AllRecoEvents_\mathrm{v}, \AllRecoEvents_\mathrm{w} )
+ \mathrm{Cov}( \AllRecoEvents_\mathrm{v}, \AllRecoEvents_\mathrm{v} )
\\
\label{eq:stat_var}
&= 0 + 0 + 0 + \mathrm{Var}( \AllRecoEvents_\mathrm{v} ) \approx
\AllRecoEvents_\mathrm{v} \,.
\end{align}

In the last approximate equality, the measured number of events
$\AllRecoEvents_\mathrm{v}$ in bin $\mathrm{v}$ is used as an estimator of its variance. Thus,
the statistical covariance between any two reconstructed bins can be estimated
simply by calculating the number of events that they have in common. This
result becomes particularly intuitive in the limits where $X$ and $Y$ are
identical and where they are disjoint.

The same derivation can also be applied to evaluate Monte Carlo statistical
uncertainties on a simulation prediction. In the case of weighted events, which
are used to adjust a nominal simulation by the NOvA~\cite{Acero2020} and
MicroBooNE~\cite{microboonegenietune} collaborations among others, the
appropriate estimator for the variance of bin $\mathrm{v}$ is the sum of the
squared weights of the events that it contains. In the limit of unit weights for
all events, this reduces to the simple event count appropriate for measured data
and an unweighted MC prediction.

\subsubsection{Unfolding complete covariances}
\label{sec:unfold_covariances}

Unlike the statistical covariances, evaluation of the systematic covariances
between bins from different kinematic distributions proceeds without any
subtleties. For the case of a MicroBooNE-style analysis, the \mbox{formula}
from Eq.~\ref{eq:CovMat} can be used unaltered for all pairs of reconstructed
bins $\recoBinIdx$ and $\secondRecoBinIdx$ to be included in the final
measurement. To avoid double-counting, however, the sum over $\trueBinIdx$ in
Eq.~\ref{eq:ExpectedRecoEvents} should include only the true bins belonging to
the same block (defined below) as reconstructed bin $\recoBinIdx$.

To simplify propagation of the results through unfolding, I recommend a
specific scheme for organizing the analysis bins. Define the term
\textit{block} to refer to a group of related true and reconstructed bins
intended for use in a measurement of the same distribution of observables
$\PhaseSpaceVec$. Apart from individual bin limits on the values of these
observables, the same signal definition should be shared by all true bins
belonging to the same block, and a single set of selection criteria should
apply to all reconstructed bins in the block. Within a properly-formed block,
there should also not be any phase-space gaps or bin overlaps; any simulated
signal interaction that fulfills all relevant event selection criteria should
belong to exactly one true bin and exactly one reconstructed bin within the
block. Note that there is no requirement for distinct blocks to have the same
signal definition, event selection criteria, or observables of interest
$\PhaseSpaceVec$. I will assume below, however, that all bins in the analysis
of interest have been grouped into valid blocks.

To proceed with the remainder of MicroBooNE-style cross-section extraction, an
unfolding matrix $\UnfoldingMatrix_\blockIdx$ should be constructed separately
for each individual block $\blockIdx$. There is no restriction on the specific
recipe to be used for computing $\UnfoldingMatrix_\blockIdx$, and a unique
unfolding method can be used in each block if desired. The formulas for
calculating the unfolding matrix, given in the preceding sections and in
Appendix~\ref{sec:Ac_wsvd}, can still be used in this case as long as they are
evaluated while ignoring all bins outside of the block of interest $\blockIdx$.
For example, the sums over true and reconstructed bins that appear in
Eqs.~\ref{eq:updated_estimator_dagostini}, \ref{eq:binprob_dagostini}, and
\ref{eq:migmat} for D'Agostini unfolding should include only bins belonging to
the current block. When computing the per-block error propagation matrix
$\ErrPropMatrix_\blockIdx$ needed for MicroBooNE-style propagation of
uncertainties (see Sec.~\ref{sec:analytic_prop}), bins outside of block
$\blockIdx$ should also be ignored in the same way.

The matrices for the individual blocks may be used to form an overall
unfolding matrix $\UnfoldingMatrix$ and an overall error propagation matrix
$\ErrPropMatrix$ via direct sums:
\begin{equation}
\label{eq:overall_unfolding_matrix}
\UnfoldingMatrix = \bigoplus_{\blockIdx = 0} \UnfoldingMatrix_\blockIdx
= \UnfoldingMatrix_0 \oplus \UnfoldingMatrix_1 \oplus \dots
= \begin{pmatrix}
\UnfoldingMatrix_0 & 0 & 0 & \dots \\
0 & \UnfoldingMatrix_1 & 0 & \dots \\
0 & 0 & \ddots & \dots \\
\vdots & \vdots & \vdots & \ddots \\
\end{pmatrix} \,,
\end{equation}
and
\begin{equation}
\label{eq:overall_err_prop_matrix}
\ErrPropMatrix = \bigoplus_{\blockIdx = 0} \ErrPropMatrix_\blockIdx \,.
\end{equation}
These can then be used to extract final cross-section results involving all
bins. Likewise, if one forms an overall detector response matrix from
the response matrices calculated for individual blocks,
\begin{equation}
\label{eq:overall_response_matrix}
\ResponseMatrix = \bigoplus_{\blockIdx = 0} \ResponseMatrix_\blockIdx \,,
\end{equation}
then Eq.~\ref{eq:add_smear_general} can be used to compute the regularization
matrix $\AddSmearMatrix$ describing the entire measurement.

In the most general case, bins belonging to different blocks may involve
different observables, and different scaling factors may be appropriate for
converting their unfolded contents to a differential cross section. I thus
update the notation from prior sections to obtain the new formulas
\begin{equation}
\label{eq:xsec_multi_block}
\left< \frac{ d^\XsecDimension\sigma }{ d\PhaseSpaceVec }
\right>_{\!\trueBinIdx}
= \frac{ \MeasuredSignalEvtCount_\trueBinIdx }
{ \IntegratedFlux_\trueBinIdx \, \NumTargets_\trueBinIdx \,
\PhaseSpaceVecWidths_\trueBinIdx } = \frac{ \sum_\recoBinIdx
\UnfoldingMatrix_{\trueBinIdx \recoBinIdx}
\, ( \AllRecoEvents_\recoBinIdx - \PredictedBkgdCount_\recoBinIdx ) }
{ \IntegratedFlux_\trueBinIdx \, \NumTargets_\trueBinIdx \,
\PhaseSpaceVecWidths_\trueBinIdx } \,,
\end{equation}
and
\begin{equation}
\label{eq:cov_multi_block}
\mathrm{Cov}\left(
\Big< \frac{d^\XsecDimension\sigma}{d\PhaseSpaceVec} \Big>_\trueBinIdx,
\Big< \frac{d^\SecondXsecDimension\sigma}{d\SecondPhaseSpaceVec}
\Big>_\secondTrueBinIdx \right)
= \frac{
\sum_{\recoBinIdx, \secondRecoBinIdx}
\ErrPropMatrix_{\trueBinIdx \recoBinIdx}
\,
\mathrm{Cov}( \AllRecoEvents_\recoBinIdx, \AllRecoEvents_\secondRecoBinIdx )
\,
\ErrPropMatrix_{\secondTrueBinIdx \secondRecoBinIdx}
}
{ \IntegratedFlux_\trueBinIdx \,
\IntegratedFlux_\secondTrueBinIdx \,
\NumTargets_\trueBinIdx \, \NumTargets_\secondTrueBinIdx \,
\PhaseSpaceVecWidths_\trueBinIdx
\, \SecondPhaseSpaceVecWidths_\secondTrueBinIdx } \,.
\end{equation}
Here $\XsecDimension$ ($\SecondXsecDimension$) is the number of observables
$\PhaseSpaceVec$ ($\SecondPhaseSpaceVec$) for which a differential cross
section is reported in true bin $\trueBinIdx$ ($\secondTrueBinIdx$), and
$\PhaseSpaceVecWidths_\trueBinIdx$
($\SecondPhaseSpaceVecWidths_\secondTrueBinIdx$) is the corresponding product
of bin widths.

True bin indices $\trueBinIdx$ and $\secondTrueBinIdx$ appear above as
subscripts on the integrated flux $\IntegratedFlux$ and number of scattering
targets $\NumTargets$. In this context, they indicate that the value of
$\IntegratedFlux$ or $\NumTargets$ that should be used is the one for the block
containing the true bin $\trueBinIdx$ or $\secondTrueBinIdx$. Differences in
the signal definition between blocks may necessitate the use of distinct values
in some cases, such as a discrepancy between the neutrino flavors of interest
($\IntegratedFlux$) or the fiducial volumes in which events are accepted
($\NumTargets$) by the analysis. All other notation in these expressions is
defined as in previous sections, and all bin indices may refer to an element of
any block.

Because the overall unfolding and error propagation matrices from
Eqs.~\ref{eq:overall_unfolding_matrix}--\ref{eq:overall_err_prop_matrix} are
block diagonal, the measured cross sections and covariances within block
$\blockIdx$ are identical to the results that would have been obtained if a
standalone measurement had been performed for block $\blockIdx$ alone using the
unfolding matrix $\UnfoldingMatrix_\blockIdx$ and error propagation matrix
$\ErrPropMatrix_\blockIdx$. However, since the reconstructed-space covariance
matrix $\mathrm{Cov}(\AllRecoEvents_\recoBinIdx,
\AllRecoEvents_\secondRecoBinIdx)$ includes complete inter-block correlations
and is propagated through the extraction procedure, these correlations are
appropriately preserved in the final result.

\subsection{Simplifying data releases}

When working with multiple blocks involving different observables
($\PhaseSpaceVec \neq \SecondPhaseSpaceVec$ above), an unfortunate feature of
the expression given in Eq.~\ref{eq:cov_multi_block} is that the units needed
to express the covariance matrix elements will vary with the true bin indices
$\trueBinIdx$ and $\secondTrueBinIdx$. In particular, it is possible for the
units of the off-diagonal elements of the covariance matrix connecting two
separate blocks to differ from the units of the covariances contained within
either block. Especially for an analysis involving many distinct blocks,
expressing the measurement in this form is cumbersome and unnecessarily
confusing.

To avoid this issue entirely, I recommend an alternative but entirely
equivalent way of representing the data. Using the same notation as in
Sec.~\ref{sec:intro}, define the flux-averaged total cross section in true bin
$\trueBinIdx$ by
\begin{equation}
\label{eq:tot_xsec_flux_avg_in_bin}
\langle \sigma \rangle_\trueBinIdx \equiv
\int_{\PhaseSpaceVec_\trueBinIdx}^{\PhaseSpaceVec_{\trueBinIdx + 1}}
\left< \frac{ d^\XsecDimension\sigma }{ d\PhaseSpaceVec } \right>
\, d\PhaseSpaceVec \,.
\end{equation}
As implied by Eq.~\ref{eq:diff_xsec_flux_avg_in_bin}, this quantity is just the
product of the average differential cross section in the bin multiplied by the
$\XsecDimension$ bin widths:
\begin{equation}
\label{sec:tot_xsec_mult}
\langle \sigma \rangle_\trueBinIdx =
\left< \frac{ d^\XsecDimension\sigma }{ d\PhaseSpaceVec }
\right>_{\!\trueBinIdx} \cdot \PhaseSpaceVecWidths_\trueBinIdx \,.
\end{equation}
It follows from this that
Eqs.~\ref{eq:xsec_multi_block}--\ref{eq:cov_multi_block} may be rewritten in
terms of the flux-averaged total cross sections as
\begin{equation}
\label{eq:tot_xsec_multi_block}
\langle \sigma \rangle_\trueBinIdx
= \frac{ \MeasuredSignalEvtCount_\trueBinIdx }
{ \IntegratedFlux_\trueBinIdx \, \NumTargets_\trueBinIdx }
= \frac{ \sum_\recoBinIdx
\UnfoldingMatrix_{\trueBinIdx \recoBinIdx}
\, ( \AllRecoEvents_\recoBinIdx - \PredictedBkgdCount_\recoBinIdx ) }
{ \IntegratedFlux_\trueBinIdx \, \NumTargets_\trueBinIdx } \,,
\end{equation}
and
\begin{equation}
\label{eq:tot_cov_multi_block}
\mathrm{Cov}\Big( \langle \sigma \rangle_\trueBinIdx,
\langle \sigma \rangle_\secondTrueBinIdx \Big)
= \frac{
\sum_{\recoBinIdx, \secondRecoBinIdx}
\ErrPropMatrix_{\trueBinIdx \recoBinIdx}
\,
\mathrm{Cov}( \AllRecoEvents_\recoBinIdx, \AllRecoEvents_\secondRecoBinIdx )
\,
\ErrPropMatrix_{\secondTrueBinIdx \secondRecoBinIdx}
}
{ \IntegratedFlux_\trueBinIdx \,
\IntegratedFlux_\secondTrueBinIdx \,
\NumTargets_\trueBinIdx \, \NumTargets_\secondTrueBinIdx } \,.
\end{equation}
Since the bin widths $\PhaseSpaceVecWidths_\trueBinIdx$ and
$\SecondPhaseSpaceVecWidths_\secondTrueBinIdx$ are exactly known by definition,
conversion to this new representation is trivial. The absence of these widths
in Eqs.~\ref{eq:tot_xsec_multi_block}--\ref{eq:tot_cov_multi_block}
conveniently allows a single set of units to be used to report all of the
measured $\langle \sigma \rangle_\trueBinIdx$ values (e.g.,
\si{\centi\meter\squared}) and their covariances (e.g.,
\si{\centi\meter\tothe{4}}).

For future cross-section analyses adopting blockwise unfolding, I therefore
recommend the following format for an associated data release. The measured
flux-averaged total cross sections $\langle \sigma \rangle_\trueBinIdx$ should
be expressed as a column vector of values ordered by bin number $\trueBinIdx$
and sharing the same units. The covariances from
Eq.~\ref{eq:tot_cov_multi_block} should likewise be expressed in a single
matrix whose rows and columns correspond to the bin indices $\trueBinIdx$ and
$\secondTrueBinIdx$, respectively. Each element of the matrix should be
expressed in units equal to the square of the units used to report $\langle
\sigma \rangle_\trueBinIdx$. If available, the regularization matrix
$\AddSmearMatrix$ should also be included with the results using the same
ordering of the bin indices. The elements of $\AddSmearMatrix$ are
dimensionless. A table listing the bin definitions and block boundaries in
order of bin number should also be provided to allow for straightforward
evaluation (by dividing by the appropriate bin widths) of differential cross
sections.

In addition to reducing the potential for unit-related confusion, this data
presentation strategy has an additional advantage. Since division by the bin
widths is performed by downstream users of the data only when needed, underflow
and overflow bins for which at least one of the widths is infinite can be
reported alongside ordinary bins without any required changes. This avoids the
need for the ad hoc treatments that are sometimes employed in the literature,
e.g., including events belonging to underflow and overflow bins in their
closest finite-width neighbors as in Refs.~\cite{tkiprd, tkiprl, uBGKI}.

\subsection{MINERvA-style measurements}
\label{sec:bu_minerva_style}

For blockwise unfolding within MINERvA's approach to cross-section extraction,
the guidelines above for organizing and presenting a measurement may be applied
largely unaltered. However, instead of building a reconstructed-space total
covariance matrix $\mathrm{Cov}(\AllRecoEvents_\recoBinIdx,
\AllRecoEvents_\secondRecoBinIdx)$ and transforming it with the error
propagation matrix $\ErrPropMatrix$, uncertainties are now quantified by
re-extracting the cross sections in multiple alternative universes as described
in Sec.~\ref{sec:multi_universe_extract}. For the statistical uncertainties,
these may include bootstrapping-based universes in which the observed bin
contents are resampled according to the appropriate underlying distribution.

Provided that the full multi-block unfolding matrix $\UnfoldingMatrix$ from
Eq.~\ref{eq:overall_unfolding_matrix} is recomputed in each universe, the
change in uncertainty quantification strategy presents no special difficulties
for the systematic variations. However, bootstrapping to evaluate the
statistical uncertainties raises a new challenge. For a MicroBooNE-like
analysis, considering correlations between individual pairs of bins is
sufficient to calculate statistical covariance matrix elements (see
Sec.~\ref{sec:stat_corr}). Construction of universes for this purpose, on the
other hand, requires statistical resampling that accounts for correlations
between all of the bins simultaneously.

To explain how this may be accomplished, I will specialize to the case of data
statistical uncertainties, i.e., those on the vector of measured event counts
$\dataRecoVec$ in each reconstructed bin. Under a bootstrapping approach, these
uncertainties are propagated by re-extracting the cross section in alternative
universes $\UnivIdx$ in which the elements $\AllRecoEvents_\recoBinIdx$ of
$\dataRecoVec$ are resampled according to Poissonian statistics to create a new
vector $\dataRecoVec^\UnivIdx$. All other quantities involved in cross-section
extraction for these universes are evaluated according to the nominal
simulation.

When events are shared between the reconstructed bins $a$, resampling the
values of each of the $\AllRecoEvents_\recoBinIdx^\UnivIdx$ according to
independent Poisson distributions will incorrectly neglect inter-bin
statistical correlations. There are two strategies that could be employed to
properly account for the correlations. One of these extends the rebinning
solution considered for two bins in Sec.~\ref{sec:stat_corr} to the general
case. Despite being mathematically correct, however, its implementation
ultimately becomes impractical. The second strategy requires manipulating
individual events rather than the bins, and it is straightforward to implement
assuming that some event-by-event information remains accessible at the
relevant stage of an analysis. For completeness, both strategies are described
below.

\subsubsection{Rebinning strategy}

For a total of $\numRecoBins$ reconstructed bins in $\dataRecoVec$ (i.e.,
included in the measurement of interest), define an alternate set of
$2^\numRecoBins - 1$ reconstructed bins that I will call the
\textit{bootstrapping basis}. The measured event counts expressed in this new
basis are contained in the vector $\BBvec$, which may be written in terms of
smaller vectors as
\begin{equation}
\BBvec = \begin{pmatrix}
\BBsubVec_{1} \\
\BBsubVec_{2} \\
\vdots \\
\BBsubVec_{\numRecoBins} \\
\end{pmatrix} \,.
\end{equation}
For $k\in\{1,2,\dots,\numRecoBins\}$, the smaller vector $\BBsubVec_{k}$ has a
total of
\begin{equation}
^{R}C_k = \binom{\numRecoBins}{k} = \frac{ \numRecoBins! }
  { k! \, (\numRecoBins - k)! }
\end{equation}
elements, each corresponding to one of the unique combinations of $k$ of the
original bins from $\dataRecoVec$. For example, each new bin represented by
$\BBsubVec_{1}$ collects events that uniquely belong to a single reconstructed
bin in $\dataRecoVec$, the bins represented by $\BBsubVec_{2}$ do the same for
events that uniquely belong to a specific pair of the original bins, and the
single element of $\BBsubVec_{\numRecoBins}$ is the number of events that are
shared by all of the original bins.

Since the definitions of all of the new bins in the bootstrapping basis are
mutually exclusive, the elements of $\BBvec$ follow independent Poisson
distributions. One may therefore form a new universe to evaluate statistical
uncertainties by resampling the elements of $\BBvec$ and then summing over
those needed to recalculate each element of $\dataRecoVec$. All statistical
correlations between the original bins will be respected automatically by
construction.

While the rebinning strategy described here might be suitable for a measurement
with a small number of original bins $R$, the exponential growth in the size of
$\BBvec$ ($2^\numRecoBins - 1$ elements) means that an eighty-bin analysis
($\numRecoBins = 80$) would have a bootstrapping basis vector $\BBvec$ with a
number of elements greater than Avogadro's number. For all realistic
situations, this will exceed the number of events in the analysis itself and
preclude the use of rebinning for generating statistical universes.

\subsubsection{Event-by-event strategy}
\label{sec:event_bootstrapping}

A practical alternative to the use of rebinning described above is a more
traditional bootstrapping technique using the individual events. Under this
approach, a new vector of reconstructed event counts $\dataRecoVec^\UnivIdx$ in
the $u$-th universe is generated by iterating over all relevant events once.
For each event, a random nonnegative integer $\xi$ is sampled from a Poisson
distribution with a mean of one. Each reconstructed bin in
$\dataRecoVec^\UnivIdx$ to which the event belongs is then incremented by
$\xi$. In the case of weighted events (which may arise, e.g., when
bootstrapping MC statistical uncertainties), the bin(s) of interest would be
incremented by $\xi \cdot w$, where $w$ is the weight of the current event.

While this method of bootstrapping requires the reconstructed bin(s) populated
by each individual event to be recorded, storing this information remains
manageable for analyses with many bins. Computational efficiency will likely be
maximized by generating all needed statistical universes
$\dataRecoVec^\UnivIdx$ using a single iteration over the events with an
independent Poisson throw in each universe $\UnivIdx$ for each event.

\subsection{T2K-style measurements}

Thanks to the flexibility of the T2K collaboration's likelihood fitting
technique, extracting blockwise-unfolded cross sections under their approach is
largely automatic once a suitable binning scheme has been defined. Either of
the two strategies for uncertainty propagation described in
Sec.~\ref{sec:t2k_unc_prop} may be applied, but see Sec.~\ref{sec:universes}
below for an important caveat. Statistical uncertainties can be handled by
creating alternate universes using the event-by-event bootstrapping treatment
described above.

\subsection{Inter-analysis covariances}

Blockwise unfolding is general enough to handle cases in which measurements in
distinct blocks are obtained from different analyses of data from the same
experiment. Nevertheless, evaluating correlated uncertainties between
cross-section results produced asynchronously raises practical difficulties
related to data preservation. To alleviate these difficulties, I consider below
the minimal information that must be available in order to calculate
inter-analysis covariances.

\subsubsection{Event lists}

Similar to blockwise unfolding within a single analysis, the major challenge
for computing statistical covariance matrix elements between separate analyses
is the need to treat the correlations arising from shared events. As discussed
previously in Secs.~\ref{sec:stat_corr}~and~\ref{sec:bu_minerva_style},
properly doing so requires knowledge of the overlaps between the contents of
all reconstructed bins from all analyses of interest. In general, this is only
feasible by tracking the reconstructed bin contents on an event-by-event basis.

The bookkeeping required for this task is ponderous but uncomplicated to
implement if it is prepared in advance. First, a unique identifier needs to be
assigned to each data and MC event considered in any of the relevant analyses.
Experimental collaborations typically do this as a matter of
course.\footnote{An example identification scheme is the set of run, subrun,
and event indices used within the art event-processing
framework~\cite{Green2012} adopted by multiple Fermilab experiments.} Second,
an event list must be prepared for each analysis. Each entry in the list must
specify a unique event identifier together with an array of zero or more
indices that represent the reconstructed bins to which the event belongs. In
the case of weighted MC events, the statistical weights should also be included
in each list entry if they are not otherwise easily accessible.

\subsubsection{Universes}
\label{sec:universes}

For systematic uncertainty quantification, the main requirement is access to
the full sets of alternative universes adopted by each of the analyses.
Situations in which these sets are inconsistent, such as when a collaboration's
approach to estimating specific systematic uncertainties has changed over time,
must necessarily be handled on a case-by-case basis. I will therefore assume
full consistency here while recognizing that careful work may be needed to
relate different descriptions of the same systematic effect across analyses.

For MicroBooNE-style measurements, the most convenient representation of the
$\UnivIdx$-th systematic universe is the vector $\recoVec^\UnivIdx$, whose
elements are the expected total event counts
$\PredictedRecoEvtCount^\UnivIdx_\recoBinIdx$ in each reconstructed bin
$\recoBinIdx$. The $\recoVec^\UnivIdx$ from each analysis can be used to
construct an overall reconstructed-space covariance matrix according to
Eq.~\ref{eq:CovMat}.

For MINERvA-style measurements, the universes are represented directly as
alternative values of the unfolded measurement
$\DiffXsecAbbrev_\trueBinIdx^\UnivIdx$ in each true bin $\trueBinIdx$. In this
form, they can easily be re-used to compute inter-analysis covariances
according to Eq.~\ref{eq:minerva_multisim}.

For T2K-style measurements, calculation of inter-analysis systematic
covariances is only well-defined under the first of the two uncertainty
propagation strategies described in Sec.~\ref{sec:t2k_unc_prop}. Under that
approach, there is a one-to-one correspondence between each alternative
universe and the extracted cross sections
$\DiffXsecAbbrev_\trueBinIdx^\UnivIdx$ obtained in its individual likelihood
fit. This allows the same universe definition $\UnivIdx$ to be used across all
analyses when computing covariances via Eq.~\ref{eq:minerva_multisim}. When
the universes are instead constructed under the second approach, i.e., based
upon the post-fit parameter covariance matrix $\mathrm{Cov}(
\fitParam_\fitParamIdx, \fitParam_\secondFitParamIdx )$ from
Eqs.~\ref{eq:t2k_hessian}--\ref{eq:t2k_param_covariance}, this is no longer
practicable. Separate analyses will, in general, have different post-fit
parameter values $\fitParamVecPF$ and covariances. Relating the distinct sets
of universes generated from these different inputs may not be achievable in
any rigorous way.

\subsubsection{Calculating the covariances}

With the event lists and universes in hand, there are two main recipes that
can be used to obtain the final inter-analysis uncertainties. When all
measurements of interest were performed in the MicroBooNE style, one may
calculate both systematic and statistical contributions to the multi-analysis
total covariance matrix in reconstructed space according to the prescriptions
from Secs.~\ref{sec:analytic_prop}~and~\ref{sec:stat_corr}. Using the overall
error propagation matrix $\ErrPropMatrix$ from
Eq.~\ref{eq:overall_err_prop_matrix}, the covariances between the final
cross-section results can then be evaluated according to
Eq.~\ref{eq:cov_multi_block}. Central values should be used for all of the
scaling factors (e.g., $\IntegratedFlux_\trueBinIdx$) that appear therein.

In the more general case where at least one analysis was performed using a
different cross-section extraction approach, a separate step is needed for the
statistical uncertainties. These must be handled by preparing alternate
universes using event-by-event bootstrapping (see
Sec.~\ref{sec:event_bootstrapping}) that populates the full set of
reconstructed bins from all of the measurements. The cross sections should then
be re-extracted in each of these universes to propagate the uncertainties. In
the re-extractions, central values should be assumed for all quantities not
subject to the relevant kind of statistical fluctuation (data, MC). The results
of this first step may be combined with the systematic universes, which are
evaluated on the unfolded measurements via Eq.~\ref{eq:minerva_multisim}, to
obtain a total inter-analysis covariance matrix.

When MicroBooNE-style measurements are included in this calculation, one must
transform the expected event counts
$\PredictedRecoEvtCount^\UnivIdx_\recoBinIdx$ used to represent each
systematic universe $\UnivIdx$ into a corresponding unfolded differential
cross section $\DiffXsecAbbrev_\trueBinIdx^\UnivIdx$. Fortunately,
Eq.~\ref{eq:cov_multi_block} implies that this amounts to the simple
expression
\begin{equation}
\DiffXsecAbbrev_\trueBinIdx^\UnivIdx
= \frac{1}{ \IntegratedFlux_\trueBinIdx \, \NumTargets_\trueBinIdx
\, \PhaseSpaceVecWidths_\trueBinIdx }
\sum_\recoBinIdx
\ErrPropMatrix_{\trueBinIdx \recoBinIdx}
\, \PredictedRecoEvtCount^\UnivIdx_\recoBinIdx \,.
\end{equation}

\section{Conditional covariance background constraint}

Scientific interest in neutrino cross-section measurements is primarily driven
by the need to refine nuclear interaction simulations\footnote{Established
neutrino scattering simulation codes for \si{GeV} energies include
GENIE~\cite{genie:2021npt, genie}, GiBUU~\cite{Mosel:2023zek, mosel:2019vhx,
buss2012}, NEUT~\cite{hayato2021, hayato2009}, and NuWro~\cite{golan2012,
golan2012b, Juszczak2006}, while ACHILLES~\cite{achilles, Isaacson:2021xty} is
an emerging effort.} and their underlying theoretical ingredients. Because
these simulations play an essential role in the design and execution of the
measurements themselves, great care must be taken to avoid biasing results
toward the interaction model used to obtain them. In appreciation of the
challenges posed by this potential bias, experimental collaborations invest
considerable effort toward quantifying relevant systematic uncertainties, and
best practices for minimizing model dependence in cross-section analyses are a
subject of ongoing discussion~\cite{Mahn2018, Betancourt2018, Avanzini:2021qlx,
nuxtract}.

Regardless of the specific methods employed in any given analysis, neutrino
cross-section extraction inherently relies upon simulation predictions in ways
that make interaction model deficiencies a significant concern. While a
thoughtful choice of the signal definition, measurement observables, and
binning scheme for an analysis can help to prevent problematic
unfolding~\cite{Avanzini:2021qlx}, mock-data studies in which a cross section
is extracted from an alternative simulation are also routinely used to detect
and diagnose bias. In analyses for which the expected background rate remains
appreciable after all event selection criteria have been applied, robust
estimation of the residual contamination becomes important. For some classes of
background, notably cosmic-ray and radiological activity, this can often be
done by means of a direct measurement performed when the neutrino beam is not
active. However, background mismodeling presents a hazard in the frequent cases
where a simulation prediction must be used.

Although details of procedures for revealing and remedying inaccurate
background estimates are often highly analysis-specific, an overall strategy is
widely adopted in the neutrino scattering literature. As a supplement to the
binning scheme used to extract the final cross-section results, one or more
sets of additional reconstructed bins, which are referred to as \textit{control
samples} or \textit{sidebands}~\cite{Mahn2018}, are defined with alternative
selection criteria intended to enhance the contribution of background events.
To provide the best possible constraints on the background prediction in the
\textit{signal region} (i.e., the original bins), the sidebands are typically
designed to cover a similar range of kinematic phase space and involve minimal
changes to the event selection. Good agreement between measured and predicted
event counts in the sideband bins is interpreted as evidence that the
simulation is reliable enough to use for background subtraction. When
significant discrepancies are found, corrective adjustments can be applied to
the simulation in various ways.

\subsection{Use of sideband constraints by experiments}
\label{sec:sidebands_exper}

A methodological strength of the T2K fitting technique for cross-section
extraction (Sec.~\ref{sec:t2k_extraction}) is that sideband-based background
constraints can be incorporated simply by including the new bins in the
likelihood fit. Correlations between the sideband bins and those used in the
final measurement will automatically be described according to the experiment's
full simulation of beam production, interaction physics, and the detector
response.

Many MINERvA cross-section analyses~\cite{MINERvA:2015jih, MINERvA:2016oql,
MINERvA:2016zyp, MINERvA:2017dzh, MINERvA:2017ipy, MINERvA:2020anu, Olivier2023,
MINERvA:2018hqn, MINERvA:2019gsf, MINERvA:2019rhx, MINERvA:2022esg,
MINERvA:2022djk, MINERvA:2023ner} share a general approach to the use of
sidebands that has also been employed in a few measurements by
T2K~\cite{T2K:2019yqu, T2K:2017qxv, T2K:2016cbz} while extracting cross sections
in the MINERvA style (Sec.~\ref{sec:minerva_extract}). To apply this method, one
may re-express the background prediction $\PredictedBkgdCount_\recoBinIdx$ in
the $\recoBinIdx$-th reconstructed bin (see Eq.~\ref{eq:extract}) as
\begin{equation}
\label{eq:scale_factor_bkgd}
\PredictedBkgdCount_\recoBinIdx = \sum_\bkgdIdx
\bkgdScaleFactor_{\bkgdIdx} \, \PredictedBkgdCount_{\recoBinIdx \bkgdIdx}
\end{equation}
where the varieties of background are indexed by the \mbox{Icelandic} letter
thorn ($\bkgdIdx$) and $\PredictedBkgdCount_{\recoBinIdx \bkgdIdx}$ is the
expected number of background events of type $\bkgdIdx$ in the $\recoBinIdx$-th
reconstructed bin. Prior to inspection of the sideband data, the scale factors
$\bkgdScaleFactor_\bkgdIdx$ are all set to unity, and
$\PredictedBkgdCount_\recoBinIdx$ thus takes a value unaltered from the nominal
simulation.

By fitting the normalization of the predicted background components $\bkgdIdx$
to the data in the sideband bins, updated values of the scale factors are
obtained. Specifics of the fitting procedure are not always exhaustively
documented and can differ substantially between analyses, including whether the
signal prediction is also allowed to vary in the fit, whether multiple
sidebands are fit separately or simultaneously, and whether the bins that will
be used for the final measurement are fit together with the sidebands.
Ultimately, the updated scale factor $\bkgdScaleFactor_\bkgdIdx$ is assigned a
value based upon the ratio of the post-fit and pre-fit predictions of the
number of background events of type $\bkgdIdx$ in the relevant sideband(s). The
updated scale factors are then applied in Eq.~\ref{eq:scale_factor_bkgd} to
compute the values of $\PredictedBkgdCount_\recoBinIdx$ actually used in
cross-section extraction.\footnote{In some analyses, the scale factors are
evaluated separately in different kinematic regions. This can be handled in the
formalism of Eq.~\ref{eq:scale_factor_bkgd} by either subdividing the
background categories $\bkgdIdx$ in terms of kinematics or by assigning the
scale factors bin-dependent values $\bkgdScaleFactor_\bkgdIdx \to
\bkgdScaleFactor_{\recoBinIdx \bkgdIdx}$ as in Ref.~\cite{Olivier2023}.}

Although rarely described explicitly~\cite{MINERvA:2016oql, T2K:2016cbz}, the
standard method of incorporating this background modeling constraint into a
MINERvA-style uncertainty treatment appears to be repeating the sideband fits
during cross-section re-extraction in each systematic universe
(Sec.~\ref{sec:multi_universe_extract}). The statistical uncertainty on the fit
results may be addressed by adding new universes in which the scale factors
themselves are varied appropriately.

While comparatively simple to incorporate into a cross-section analysis, the
most straightforward implementation of the MINERvA background constraint
strategy has some limitations relative to the T2K likelihood fit. First, since
the MINERvA method adjusts only normalization scale factors, sideband
constraints are by construction unable to influence the shape of the individual
classes of background to be subtracted. Second, in cases where the original
bins to be used for the cross-section measurement are not included in the scale
factor fit, a 100\% correlation is implicitly assumed to exist between the
background normalization in the sidebands and in the signal region. Obvious
problems arising from these limitations can be detected by examining the level
of agreement between the constrained simulation and data. If necessary,
modifications can be made to improve the agreement, such as assigning separate
scale factors in different kinematic ranges.

In contrast to T2K and MINERvA, MicroBooNE has not reported the use of control
samples to validate the background model or apply a data-driven adjustment to
it for most of their differential cross section measurements released so
far~\cite{mcc8CCincl, mcc8ccnppaper, Abratenko:2020acr, tkiprl, tkiprd,
nueCCDiffXSec, WireCellCCinclPRL, BNBnueXsec, ub2p, microboone:2023foc, uBGKI,
MicroBooNE:2024yxe, MicroBooNE:2024zwf, MicroBooNE:2024zkh}. The predicted
background from simulation is simply used as-is without any analysis-specific
changes to the central value or the modeling uncertainties. Such an approach
has occasionally also been used by MINERvA~\cite{MINERvA:2020zzv,
MINERvA:2021owq, MINERvA:2021wjs}, but only in cases where the expected
background is very small.

Two of MicroBooNE's most recent measurements of differential cross sections,
which examine neutral-current production of neutral
pions~\cite{MicroBooNE:2024sec} and mesonless charged-current interactions with
protons in the final state~\cite{MicroBooNE:2024tmp}, each include supplemental
material that documents checks of the background model against sideband data.
While these checks provide evidence for the adequacy of the MicroBooNE
simulation's background prediction, no attempt is made to improve it in light
of the sidebands during the cross-section extraction procedure.

A possible reason for the omission of a sideband-based background constraint in
these analyses by MicroBooNE is that it may not necessarily be obvious how to
incorporate such a constraint into the systematics treatment needed to apply
the Wiener-SVD unfolding method (Sec.~\ref{sec:analytic_prop}). However,
multiple MicroBooNE measurements of flux-averaged \textit{total} cross sections
have included the use of sidebands~\cite{uBEta, MicroBooNE:2022cls,
MicroBooNE:2022zhr}. One of these, a measurement of $\eta$ meson
production~\cite{uBEta}, used a unique background constraint method that
appears to be new to the neutrino cross-section literature. In the remainder of
this section, I adapt and generalize this procedure to provide a recipe for
incorporating sideband-based refinements to background estimation in
MicroBooNE-style measurements of unfolded differential cross sections. I call
the generalized technique the \textit{conditional covariance background
constraint} (CCBC).

It should be noted that the CCBC approach presented here shares many
mathematical similarities to the \textit{data-driven model validation} (DDMV)
procedure introduced by MicroBooNE as a means of stringently testing the
suitability of their simulation for cross-section extraction in specific
observables~\cite{MicroBooNE:2024zwf}. While the two techniques share the same
mathematical formalism for evaluating conditional covariances, the application
of DDMV tends to focus on unfolding rather than background estimation, and its
purpose is to verify that a given input model is sufficiently reliable rather
than to prepare a modified version of that model to use for background
subtraction.

\subsection{Conditional covariance formalism}
\label{sec:cc_formalism}

The mathematical foundation for the CCBC is the conditional multivariate
Gaussian distribution. Let the random vector $\randVec$ follow a multivariate
Gaussian distribution with mean vector $\meanVec$ and covariance matrix
$\randVecCovMat$. Without loss of generality, choose the ordering of the
elements of $\randVec$ such that it can be partitioned into two smaller random
vectors
\begin{equation}
\randVec = \begin{pmatrix} \randVec_\randVecFirstIdx
\\ \randVec_\randVecSecondIdx \end{pmatrix} \,.
\end{equation}
The mean vector and covariance matrix can thus be written in the form
\begin{equation}
\meanVec = \begin{pmatrix} \meanVec_\randVecFirstIdx
\\ \meanVec_\randVecSecondIdx \end{pmatrix} \,,
\end{equation}
and
\begin{equation}
\randVecCovMat = \begin{pmatrix}
\randVecCovMat_{\randVecFirstIdx \randVecFirstIdx}
& \randVecCovMat_{\randVecFirstIdx \randVecSecondIdx} \\
\randVecCovMat_{\randVecSecondIdx \randVecFirstIdx}
& \randVecCovMat_{\randVecSecondIdx \randVecSecondIdx} \\
\end{pmatrix}
\end{equation}
respectively. By construction, $\randVecCovMat_{\randVecFirstIdx
\randVecSecondIdx} = \randVecCovMat_{\randVecSecondIdx \randVecFirstIdx}^{T}$,
and $\randVec_\randVecFirstIdx$ ($\randVec_\randVecSecondIdx$) follows a
multivariate Gaussian distribution with mean $\meanVec_\randVecFirstIdx$
($\meanVec_\randVecSecondIdx$) and covariance matrix
$\randVecCovMat_{\randVecFirstIdx \randVecFirstIdx}$
($\randVecCovMat_{\randVecSecondIdx \randVecSecondIdx}$). Given the observation
that the elements of $\randVec_\randVecSecondIdx$ have some definite values
$\randVecVal_\randVecSecondIdx$, the conditional probability distribution
$P(\randVec_\randVecFirstIdx \,|\, \randVec_\randVecSecondIdx =
\randVecVal_\randVecSecondIdx)$ of $\randVec_\randVecFirstIdx$ is multivariate
Gaussian with mean
\begin{equation}
\label{eq:cond_mean}
\meanVec_\randVecFirstIdx^\randVecCondLabel
= \meanVec_\randVecFirstIdx
+ \randVecCovMat_{\randVecFirstIdx \randVecSecondIdx}
\cdot \randVecCovMat_{\randVecSecondIdx \randVecSecondIdx}^{-1}
\cdot (\randVecVal_\randVecSecondIdx - \meanVec_\randVecSecondIdx) \,,
\end{equation}
and covariance matrix
\begin{equation}
\label{eq:cond_cov}
\randVecCovMat_{\randVecFirstIdx \randVecFirstIdx}^\randVecCondLabel
= \randVecCovMat_{\randVecFirstIdx \randVecFirstIdx}
- \randVecCovMat_{\randVecFirstIdx \randVecSecondIdx}
\cdot \randVecCovMat_{\randVecSecondIdx \randVecSecondIdx}^{-1}
\cdot \randVecCovMat_{\randVecSecondIdx \randVecFirstIdx} \,.
\end{equation}
A proof in which the results from Eqs.~\ref{eq:cond_mean}--\ref{eq:cond_cov}
are derived is given in Ref.~\cite{Morris2007}.

\subsection{Application to low-energy excess analyses}

A primary motivation for the MicroBooNE experiment was the observation by
LSND~\cite{LSNDexcess1,LSNDexcess2,LSNDexcess3,LSNDexcess4} and
MiniBooNE~\cite{MiniBooNEExcessOld, MiniBooNEExcess} of an anomalous excess of
electron (anti)neutrino candidate events at low energies. In all of
MicroBooNE's first analyses investigating possible explanations for the excess,
including true $\nu_e$ appearance~\cite{MicroBooNE:2021tya, MicroBooNE:2021wad,
MicroBooNE:2021pvo, MicroBooNE:2021nxr} and an unexpectedly large rate of
neutral-current $\Delta$ baryon production followed by radiative
decay~\cite{MicroBooNE:2021zai}, the conditional covariance formalism from the
previous subsection played a major role. Since the presence or absence of an
LSND/MiniBooNE-like low-energy excess could only be judged relative to an
expectation derived from simulation, robust predictions for the selected event
rates were an essential ingredient for interpretation of the measurements.

MicroBooNE's strategy for bolstering confidence in their simulation predictions
involved both global improvements adopted for general collaboration
use\footnote{An example is the ``MicroBooNE Tune''~\cite{microboonegenietune}
of the GENIE neutrino event generator.} as well as analysis-specific model
constraints. Each of the specific constraints assumed that the expected event
counts in the signal region and one or more control samples jointly followed a
multivariate Gaussian distribution. The mean of this distribution was taken to
be the central-value prediction of the nominal simulation, and the systematic
portion of the covariance matrix was evaluated as in Eq.~\ref{eq:CovMat},
except that there was no special treatment (see
Eq.~\ref{eq:ExpectedRecoEvents}) for the interaction model uncertainties. By
conditioning on the measured event counts in each of the control sample bins
and applying Eqs.~\ref{eq:cond_mean}--\ref{eq:cond_cov}, a refined
signal-region prediction with a smaller uncertainty was obtained and used to
evaluate the final results.

\subsection{Application to $\pmb{\eta}$ production measurement}

The MicroBooNE analysis reported in Ref.~\cite{uBEta} measured a flux-averaged
total cross section for neutrino-induced $\eta$ production. Since the event
selection relied upon identification of photon pairs with kinematics consistent
with the decay $\eta \to \gamma\gamma$, photons arising from neutral pion
decays represented the dominant source of background. To provide a data-driven
constraint of this background, the single bin used to accumulate candidate
signal events was supplemented with two control sample bins. The first (second)
control sample bin used a modified selection to enhance the contribution from
events containing exactly one (two or more) $\pi^0$ in the final state.

Inspired by the conditional covariance technique used in the low-energy excess
analyses, a similar approach was used to obtain updated estimates for the
single- and multi-$\pi^0$ background contributions in the signal region. In
this case, each of these two dominant backgrounds is constrained individually
using the appropriate control sample bin. Let
$\PredictedBkgdCount_\bkgdIdx^\signalRegionLabel$ denote the expected number of
background events of type $\bkgdIdx$ (either single- or multi-$\pi^0$) in the
signal region bin. The constrained prediction for this quantity used in
Ref.~\cite{uBEta} was calculated via an expression equivalent to
\begin{equation}
\label{eq:eta_mean}
\PredictedBkgdCount_\bkgdIdx^{\signalRegionLabel,\randVecCondLabel}
= \PredictedBkgdCount_\bkgdIdx^{\signalRegionLabel}
+ \frac{ \mathrm{Cov}( \PredictedBkgdCount_\bkgdIdx^{\signalRegionLabel},
\PredictedBkgdCount_\bkgdIdx^{\controlSampleLabel} ) }
{ \mathrm{Var}( \PredictedBkgdCount_\bkgdIdx^{\controlSampleLabel} ) }
\cdot ( \AllRecoEvents^\controlSampleLabel
- \PredictedRecoEvtCount^\controlSampleLabel ) \,.
\end{equation}
In the control sample bin of interest,
$\PredictedBkgdCount_\bkgdIdx^\controlSampleLabel$ is the expected number of
background events of type $\bkgdIdx$, $\AllRecoEvents^\controlSampleLabel$ is
the measured total number of events, and
$\PredictedRecoEvtCount^\controlSampleLabel$ is the total number of events of
all kinds predicted by the nominal simulation. The covariance $\mathrm{Cov}(
\PredictedBkgdCount_\bkgdIdx^{\signalRegionLabel},
\PredictedBkgdCount_\bkgdIdx^{\controlSampleLabel} )$ and variance
\begin{equation}
\mathrm{Var}( \PredictedBkgdCount_\bkgdIdx^{\controlSampleLabel} )
= \mathrm{Cov}( \PredictedBkgdCount_\bkgdIdx^{\controlSampleLabel},
\PredictedBkgdCount_\bkgdIdx^{\controlSampleLabel} )
\end{equation}
were calculated using an expression similar to Eq.~\ref{eq:CovMat} except that
only background events of type $\bkgdIdx$ were included rather than all events.

By analogy with Eq.~\ref{eq:cond_cov}, the uncertainty on the constrained
background prediction in the signal region was represented by the variance
\begin{equation}
\label{eq:eta_var}
\mathrm{Var}(
\PredictedBkgdCount_\bkgdIdx^{\signalRegionLabel,\randVecCondLabel} )
= \mathrm{Var}( \PredictedBkgdCount_\bkgdIdx^{\signalRegionLabel} )
- \frac{ \big[ \mathrm{Cov}( \PredictedBkgdCount_\bkgdIdx^{\signalRegionLabel},
\PredictedBkgdCount_\bkgdIdx^{\controlSampleLabel} ) \big]^2 }
{ \mathrm{Var}( \PredictedBkgdCount_\bkgdIdx^{\controlSampleLabel} ) } \,.
\end{equation}
This procedure was repeated twice to obtain updated predictions and
uncertainties for both the single- and multi-$\pi^0$ backgrounds in the signal
region.

Although innovative as a first application of the conditional covariance
formalism to a sideband-based background constraint for a neutrino
cross-section measurement, the approach adopted for the MicroBooNE $\eta$
production result has some limitations that must be addressed to create a more
general technique. First, the expressions given in
Eqs.~\ref{eq:eta_mean}~and~\ref{eq:eta_var} are appropriate only for a
single-bin analysis. However, adapting them for a multi-bin measurement is
straightforward in light of the matrix notation used in
Eqs.~\ref{eq:cond_mean}~and~\ref{eq:cond_cov}. Second, the $\eta$ analysis
considers each class of background $\bkgdIdx$ in isolation from the others and
the signal, which neglects potentially important relationships between event
categories in the general case. Third, while the present approach provides a
recipe for updating the background prediction and its uncertainty in the signal
region, the Wiener-SVD unfolding technique favored by MicroBooNE requires an
input covariance matrix describing the uncertainty on the \mbox{\textit{total}}
event counts (both signal and background) in each reconstructed bin. Due to
systematic uncertainties that may be strongly correlated between signal and
background events (e.g., those due to the neutrino flux model), one cannot
simply assume that the constraint procedure will leave the signal contribution
to the total covariance matrix unchanged. The CCBC procedure described below
removes all three of these limitations.

\subsection{Assumption of joint Gaussian distribution}
\label{sec:assume_joint_Gaussian}

In the MicroBooNE approach to cross-section extraction, a covariance matrix
$\mathrm{Cov}( \PredictedRecoEvtCount_\recoBinIdx,
\PredictedRecoEvtCount_\secondRecoBinIdx )$ is constructed that describes the
uncertainty on the expected event counts $\PredictedRecoEvtCount_\recoBinIdx$
in each reconstructed bin $\recoBinIdx$ (Sec.~\ref{sec:analytic_prop}). The
multiple-universe expression (Eq.~\ref{eq:CovMat}) used to evaluate the
covariance matrix elements is based on the assumption that the
$\PredictedRecoEvtCount_\recoBinIdx$ jointly follow a multivariate Gaussian
distribution with the mean values given by the central-value prediction
$\PredictedRecoEvtCount_\recoBinIdx^\mathrm{CV}$ from the nominal simulation.
This description is also compatible with a somewhat more stringent assumption
that will form the basis of the CCBC.

Divide the expected event counts into signal
($\PredictedSignalEvtCount_\recoBinIdx$) and background
($\PredictedBkgdCount_\recoBinIdx$) contributions:
\begin{equation}
\label{eq:divided_evt_counts}
\PredictedRecoEvtCount_\recoBinIdx = \PredictedSignalEvtCount_\recoBinIdx
+ \PredictedBkgdCount_\recoBinIdx \,.
\end{equation}
The covariance matrix element can thus also be subdivided according to
\begin{align}
\nonumber
\mathrm{Cov}( \PredictedRecoEvtCount_\recoBinIdx,
\PredictedRecoEvtCount_\secondRecoBinIdx )
& =
\mathrm{Cov}(  \PredictedSignalEvtCount_\recoBinIdx,
\PredictedSignalEvtCount_\secondRecoBinIdx )
+
\mathrm{Cov}( \PredictedSignalEvtCount_\recoBinIdx,
\PredictedBkgdCount_\secondRecoBinIdx )
\\ \label{eq:divided_cov} &+
\mathrm{Cov}( \PredictedBkgdCount_\recoBinIdx,
\PredictedSignalEvtCount_\secondRecoBinIdx )
+
\mathrm{Cov}( \PredictedBkgdCount_\recoBinIdx,
\PredictedBkgdCount_\secondRecoBinIdx ) \,.
\end{align}
By substituting the right-hand side of Eq.~\ref{eq:divided_evt_counts} into
Eq.~\ref{eq:CovMat}, one may obtain separate multiple-universe expressions for
each term on the right-hand side of Eq.~\ref{eq:divided_cov}. For example,
\begin{equation}
\mathrm{Cov}( \PredictedSignalEvtCount_\recoBinIdx,
\PredictedBkgdCount_\secondRecoBinIdx )
= \frac{ 1 }{ \UnivCount } \sum_{\UnivIdx = 1}^{\UnivCount} \big(
\PredictedSignalEvtCount_{\recoBinIdx}^{\UnivIdx} -
\PredictedSignalEvtCount_{\recoBinIdx}^\mathrm{CV} \big)
\big( \PredictedBkgdCount_{\secondRecoBinIdx}^{\UnivIdx}
- \PredictedBkgdCount_{\secondRecoBinIdx}^\mathrm{CV} \big) \,.
\end{equation}

To follow the MicroBooNE prescription for handling neutrino interaction model
uncertainties, the signal prediction in the $\UnivIdx$-th alternate universe
$\PredictedSignalEvtCount_{\recoBinIdx}^{\UnivIdx}$ should be computed as in
Eq.~\ref{eq:ExpectedRecoEvents}. That is, for any alternate universe $\UnivIdx$
in which the neutrino interaction model is varied,
\begin{equation}
\PredictedSignalEvtCount_{\recoBinIdx}^{\UnivIdx}
= \sum_{\trueBinIdx} \ResponseMatrix_{\recoBinIdx \trueBinIdx}^{\UnivIdx}
\, \PredictedSignalEvtCount^\mathrm{CV}_\trueBinIdx \,,
\end{equation}
where the response matrix elements $\ResponseMatrix_{\recoBinIdx
\trueBinIdx}^{\UnivIdx}$ are evaluated in the $\UnivIdx$-th universe,
$\PredictedSignalEvtCount^\mathrm{CV}_\trueBinIdx$ is the central-value
prediction for the number of signal events in the $\trueBinIdx$-th true bin,
and the sum runs over all true bins in the block of interest (see
Sec.~\ref{sec:unfold_covariances}).

As demonstrated above, the total covariance matrix $\mathrm{Cov}(
\PredictedRecoEvtCount_\recoBinIdx, \PredictedRecoEvtCount_\secondRecoBinIdx )$
is separable into components describing the uncertainty on the signal, the
background, and their covariances with each other. The individual components
are calculable using essentially the same procedure as the combined matrix,
which is based on an assumption of joint Gaussianity of the
$\PredictedRecoEvtCount_\recoBinIdx$. I will therefore assume that the signal
and background event counts (as opposed to just their sums
$\PredictedRecoEvtCount_\recoBinIdx$) are jointly multivariate Gaussian random
variables with means given by the central-value predictions
$\PredictedSignalEvtCount^\mathrm{CV}_\recoBinIdx$ and
$\PredictedBkgdCount^\mathrm{CV}_\recoBinIdx$, respectively.

\subsection{Generalized constraint}

To adapt the background constraint strategy from Ref.~\cite{uBEta} to be
suitable for MicroBooNE-style cross-section extraction, updated predictions for
both the expected background $\PredictedBkgdCount_\recoBinIdx$ and the total
covariance matrix $\mathrm{Cov}( \PredictedRecoEvtCount_\recoBinIdx,
\PredictedRecoEvtCount_\secondRecoBinIdx )$ must be obtained from the
procedure. This can be accomplished by simultaneously conditioning the
signal-only ($\sigVec_\signalRegionLabel$) and the background-only
($\bkgdVec_\signalRegionLabel$) vectors of predicted event counts in the signal
region on the measurement ($\dataRecoVec_\controlSampleLabel$) in the control
sample bins.

\subsubsection{Sideband data statistical uncertainty}
\label{sec:sideband_data_stat_unc}

A subtlety that arises when applying the result from
Eq.~\ref{eq:cond_mean} is that $\randVecVal_\randVecSecondIdx$ is treated as
exactly known, while $\dataRecoVec_\controlSampleLabel$ is subject to data
statistical uncertainty. One may easily account for this by assigning the extra
uncertainty to the model prediction. Let
\begin{equation}
\recoVec_\controlSampleLabel = \sigVec_\controlSampleLabel
+ \bkgdVec_\controlSampleLabel
\end{equation}
denote the vector of expected total event counts in the control sample bins,
with $\sigVec_\controlSampleLabel$ ($\bkgdVec_\controlSampleLabel$)
representing the contribution of signal (background) events. As a consequence
of the joint Gaussianity assumption discussed above,
$\recoVec_\controlSampleLabel$ is itself a Gaussian random vector
that can be described in terms of a mean given
by the central-value predictions from simulation
\begin{equation}
\recoVec_\controlSampleLabel^\mathrm{CV}
  = \sigVec_\controlSampleLabel^\mathrm{CV}
+ \bkgdVec_\controlSampleLabel^\mathrm{CV} \,,
\end{equation}
and a covariance matrix that may be written in terms of
the signal and background contributions as
\begin{equation}
\label{eq:nC_cov}
\auxCovMat_{\recoVec_\controlSampleLabel \recoVec_\controlSampleLabel}
= \auxCovMat_{\sigVec_\controlSampleLabel \sigVec_\controlSampleLabel}
+ \auxCovMat_{\sigVec_\controlSampleLabel \bkgdVec_\controlSampleLabel}
+ \auxCovMat_{\bkgdVec_\controlSampleLabel \sigVec_\controlSampleLabel}
+ \auxCovMat_{\bkgdVec_\controlSampleLabel \bkgdVec_\controlSampleLabel} \,.
\end{equation}
Systematic contributions to each of the component covariance matrices in
Eq.~\ref{eq:nC_cov} are evaluated according to the generalized
multiple-universe procedure discussed in the previous subsection. Statistical
covariances should be calculated accounting for shared events as described in
Sec.~\ref{sec:stat_corr}.

As presented above, $\auxCovMat_{\recoVec_\controlSampleLabel
\recoVec_\controlSampleLabel}$ represents only the total uncertainty (both
statistical and systematic) on the model prediction
$\recoVec_\controlSampleLabel$. To account for the data statistical
uncertainty on the measurement $\dataRecoVec_\controlSampleLabel$
when applying the conditional constraint, define a new random vector
\begin{equation}
\recoWithUncVec_\controlSampleLabel \equiv \recoVec_\controlSampleLabel
+ \pmb{\delta}_\controlSampleLabel \,,
\end{equation}
where $\pmb{\delta}_\controlSampleLabel$ is multivariate Gaussian
with a mean of zero and a covariance matrix
$\auxCovMat_{\dataRecoVec_\controlSampleLabel
  \dataRecoVec_\controlSampleLabel}$ that corresponds to the data statistical
uncertainty on $\dataRecoVec_\controlSampleLabel$.
I will assume here and elsewhere that the predicted event counts before the
constraint are uncorrelated with the measured event counts in both the control
sample bins and the signal region bins. Under this assumption, the covariance
matrix for the new random vector is
\begin{equation}
\auxCovMat_{\recoWithUncVec_\controlSampleLabel
  \recoWithUncVec_\controlSampleLabel}
= \auxCovMat_{\recoVec_\controlSampleLabel \recoVec_\controlSampleLabel}
+ \auxCovMat_{\dataRecoVec_\controlSampleLabel
  \dataRecoVec_\controlSampleLabel} \,.
\end{equation}
Since the data statistical uncertainty is already included in the covariance
matrix $\auxCovMat_{\recoWithUncVec_\controlSampleLabel
  \recoWithUncVec_\controlSampleLabel}$, one may now treat
$\dataRecoVec_\controlSampleLabel$ as exactly known when it is compared to the
prediction $\recoWithUncVec_\controlSampleLabel$.

\subsubsection{Applying the constraint}
\label{sec:appl_constr}

From the results above and in Sec.~\ref{sec:cc_formalism},
one may condition the background prediction in the signal region
$\bkgdVec_\signalRegionLabel$ on the measured total event counts
$\dataRecoVec_\controlSampleLabel$ in the control sample bins to obtain
a constrained background prediction
$\bkgdVec_{\signalRegionLabel}^{\randVecCondLabel}$ with mean
\begin{equation}
\label{eq:ccbc_bkgd}
\bkgdVec_{\signalRegionLabel}^{\randVecCondLabel,\mathrm{CV}} =
\bkgdVec_{\signalRegionLabel}^\mathrm{CV}
 + \auxCovMat_{\bkgdVec_\signalRegionLabel \recoVec_\controlSampleLabel}
\cdot
\auxCovMat_{\recoWithUncVec_\controlSampleLabel
  \recoWithUncVec_\controlSampleLabel}^{-1}
\cdot \big( \dataRecoVec_\controlSampleLabel
- \recoVec_{\controlSampleLabel}^\mathrm{CV} \big)
\end{equation}
and covariance matrix
\begin{equation}
\label{eq:ccbc_bkgd_cov}
\auxCovMat_{ \bkgdVec_{\signalRegionLabel}
  \bkgdVec_{\signalRegionLabel} }^\randVecCondLabel
= \auxCovMat_{ \bkgdVec_{\signalRegionLabel} \bkgdVec_{\signalRegionLabel} }
- \auxCovMat_{\bkgdVec_{\signalRegionLabel}
\recoVec_\controlSampleLabel}
  \cdot \auxCovMat_{\recoWithUncVec_\controlSampleLabel
    \recoWithUncVec_\controlSampleLabel}^{-1} \cdot
\auxCovMat^{T}_{ \bkgdVec_{\signalRegionLabel} \recoVec_\controlSampleLabel }
\,.
\end{equation}
Simplifying to the case of a single background category $\bkgdIdx$ and single
bins for both the signal region and the control sample, the expression in
Eq.~\ref{eq:ccbc_bkgd} only becomes equivalent to Eq.~\ref{eq:eta_mean} under
the approximation that the contributions of the signal and non-$\bkgdIdx$
backgrounds to the control sample are either negligible or exactly known. The
more general expression above thus enables constraints from sidebands in which
the background(s) of interest are imperfectly isolated from other kinds of
events.

One may similarly apply this procedure to the signal prediction
$\sigVec_\signalRegionLabel$ to obtain a constrained version
$\sigVec_\signalRegionLabel^\randVecCondLabel$ with mean
\begin{equation}
\label{eq:ccbc_sig}
\sigVec_{\signalRegionLabel}^{\randVecCondLabel,\mathrm{CV}} =
\sigVec_{\signalRegionLabel}^\mathrm{CV}
 + \auxCovMat_{\sigVec_\signalRegionLabel \recoVec_\controlSampleLabel}
\cdot
\auxCovMat_{\recoWithUncVec_\controlSampleLabel
  \recoWithUncVec_\controlSampleLabel}^{-1}
\cdot \big( \dataRecoVec_\controlSampleLabel
- \recoVec_{\controlSampleLabel}^\mathrm{CV} \big)
\end{equation}
and covariance matrix
\begin{equation}
\label{eq:ccbc_cov_sig}
\auxCovMat_{ \sigVec_{\signalRegionLabel}
  \sigVec_{\signalRegionLabel} }^\randVecCondLabel
= \auxCovMat_{ \sigVec_{\signalRegionLabel} \sigVec_{\signalRegionLabel} }
- \auxCovMat_{\sigVec_{\signalRegionLabel}
\recoVec_\controlSampleLabel}
  \cdot \auxCovMat_{\recoWithUncVec_\controlSampleLabel
    \recoWithUncVec_\controlSampleLabel}^{-1} \cdot
\auxCovMat^{T}_{ \sigVec_{\signalRegionLabel} \recoVec_\controlSampleLabel }
\,.
\end{equation}
The covariance between the two constrained event counts is then given by
\begin{align}
\nonumber
\auxCovMat_{ \sigVec_{\signalRegionLabel}
  \bkgdVec_{\signalRegionLabel} }^\randVecCondLabel
&= \mathrm{Cov}\big(
\sigVec_{\signalRegionLabel}^\randVecCondLabel,
\bkgdVec_{\signalRegionLabel}^\randVecCondLabel
\big) = \big( \auxCovMat_{ \bkgdVec_{\signalRegionLabel}
  \sigVec_{\signalRegionLabel} }^\randVecCondLabel \big)^{T} \\
&= \label{eq:ccbc_cov_sig_bkgd}
\auxCovMat_{ \sigVec_{\signalRegionLabel} \bkgdVec_{\signalRegionLabel} }
- \auxCovMat_{\sigVec_{\signalRegionLabel}
\recoVec_\controlSampleLabel}
  \cdot \auxCovMat_{\recoWithUncVec_\controlSampleLabel
    \recoWithUncVec_\controlSampleLabel}^{-1} \cdot
\auxCovMat^{T}_{ \bkgdVec_{\signalRegionLabel} \recoVec_\controlSampleLabel }
\,.
\end{align}
The constrained prediction for the total event counts in the signal region
$\recoVec_\signalRegionLabel^\randVecCondLabel$ then has mean
\begin{equation}
\recoVec_{\signalRegionLabel}^{\randVecCondLabel,\mathrm{CV}}
= \sigVec_{\signalRegionLabel}^{\randVecCondLabel,\mathrm{CV}}
+ \bkgdVec_{\signalRegionLabel}^{\randVecCondLabel,\mathrm{CV}}
\end{equation}
and covariance matrix
\begin{align}
\nonumber
\auxCovMat_{ \recoVec_{\signalRegionLabel}
  \recoVec_{\signalRegionLabel} }^\randVecCondLabel &=
\mathrm{Cov}\big(
  \recoVec_{\signalRegionLabel}^\randVecCondLabel,
  \recoVec_{\signalRegionLabel}^\randVecCondLabel \big) \\
&=
\auxCovMat_{ \sigVec_{\signalRegionLabel}
  \sigVec_{\signalRegionLabel} }^\randVecCondLabel
+ \auxCovMat_{ \sigVec_{\signalRegionLabel}
  \bkgdVec_{\signalRegionLabel} }^\randVecCondLabel
+ \auxCovMat_{ \bkgdVec_{\signalRegionLabel}
  \sigVec_{\signalRegionLabel} }^\randVecCondLabel
+ \auxCovMat_{ \bkgdVec_{\signalRegionLabel}
  \bkgdVec_{\signalRegionLabel} }^\randVecCondLabel \,.
\end{align}

\subsubsection{Background-only constraint}
\label{sec:constraint_bkgd_only}

Although the constrained signal prediction
$\sigVec_\signalRegionLabel^\randVecCondLabel$ obtained in the previous
subsection is mathematically well-defined, its use in cross-section extraction
is not always advisable. For control sample(s) that are designed to constrain
relevant backgrounds, a signal constraint based upon them may introduce a high
degree of model dependence into the final results; one must rely on simulation
to accurately relate a background prediction $\bkgdVec_\controlSampleLabel$ in
a background-dominated region $\controlSampleLabel$ to a signal prediction
$\sigVec_\signalRegionLabel$ in a signal-dominated region $\signalRegionLabel$
of phase space.

While the original constraint described in Sec.~\ref{sec:appl_constr} may be
suitable for some analyses in which the control region $\controlSampleLabel$
contains at least some signal-rich bins, my recommended approach involves more
conservatively applying the constraint to the background prediction
$\bkgdVec_\signalRegionLabel$ only. This yields the same mean and covariance
matrix for $\bkgdVec_\signalRegionLabel^\randVecCondLabel$ from
Eqs.~\ref{eq:ccbc_bkgd}--\ref{eq:ccbc_bkgd_cov}. The signal prediction
$\sigVec_\signalRegionLabel$ is left unaltered, leading to a new constrained
total event count prediction $\recoBkgdConstVec_\signalRegionLabel$ with mean
\begin{equation}
\label{sec:tot_constr_B}
\recoBkgdConstVec_\signalRegionLabel^\mathrm{CV}
  = \sigVec_\signalRegionLabel^\mathrm{CV}
  + \bkgdVec_\signalRegionLabel^{\randVecCondLabel,\mathrm{CV}}
\end{equation}
and covariance matrix
\begin{align}
\nonumber
\auxCovMat_{ \recoBkgdConstVec_{\signalRegionLabel}
  \recoBkgdConstVec_{\signalRegionLabel} } &=
\mathrm{Cov}\big(
  \recoBkgdConstVec_{\signalRegionLabel},
  \recoBkgdConstVec_{\signalRegionLabel} \big) \\
&= \nonumber
\auxCovMat_{ \sigVec_{\signalRegionLabel}
  \sigVec_{\signalRegionLabel} }
+ \auxCovMat_{ \sigVec_{\signalRegionLabel}
  \bkgdVec_{\signalRegionLabel} }^\randVecCondLabel
+ \auxCovMat_{ \bkgdVec_{\signalRegionLabel}
  \sigVec_{\signalRegionLabel} }^\randVecCondLabel
+ \auxCovMat_{ \bkgdVec_{\signalRegionLabel}
  \bkgdVec_{\signalRegionLabel} }^\randVecCondLabel \\[1mm]
&= \label{sec:tot_constr_B_cov}
\auxCovMat_{ \recoVec_{\signalRegionLabel}
  \recoVec_{\signalRegionLabel} }^\randVecCondLabel
- \auxCovMat_{ \sigVec_{\signalRegionLabel}
  \sigVec_{\signalRegionLabel} }^\randVecCondLabel
+ \auxCovMat_{ \sigVec_{\signalRegionLabel}
  \sigVec_{\signalRegionLabel} } \,.
\end{align}
In deriving the second line of Eq.~\ref{sec:tot_constr_B_cov}, I have
noted that the constrained covariance between $\sigVec_\signalRegionLabel$
and $\bkgdVec_\signalRegionLabel$ is the same whether the constraint is
applied to either or both of these vectors, i.e.,
\begin{align}
\nonumber
\auxCovMat_{ \sigVec_{\signalRegionLabel}
  \bkgdVec_{\signalRegionLabel} }^\randVecCondLabel
&= \mathrm{Cov}\big( \sigVec_\signalRegionLabel^\randVecCondLabel,
  \bkgdVec_{\signalRegionLabel}^\randVecCondLabel \big)
= \mathrm{Cov}\big( \sigVec_\signalRegionLabel,
  \bkgdVec_{\signalRegionLabel}^\randVecCondLabel \big) \\
&= \mathrm{Cov}\big( \sigVec_\signalRegionLabel^\randVecCondLabel,
  \bkgdVec_{\signalRegionLabel} \big)
= \big( \auxCovMat_{ \bkgdVec_{\signalRegionLabel}
  \sigVec_{\signalRegionLabel} }^\randVecCondLabel \big)^{T} \,.
\end{align}
This result follows from the definitions above, the symmetric nature of
covariance matrices, and elementary rules of linear algebra.

Before proceeding with cross-section extraction itself, one may check that the
updated signal-region prediction
$\recoBkgdConstVec_\signalRegionLabel$ provides a good
description of the measured reconstructed event counts
$\dataRecoVec_\signalRegionLabel$ by computing the post-constraint chi-squared
statistic
\begin{equation}
\chi^2_\mathrm{post} = \big( \dataRecoVec_\signalRegionLabel
- \recoBkgdConstVec_{\signalRegionLabel}^\mathrm{CV}
\big)^T \cdot \auxCovMat_{ \recoBkgdConstWithUncVec_\signalRegionLabel
  \recoBkgdConstWithUncVec_\signalRegionLabel }^{-1} \cdot
\big( \dataRecoVec_\signalRegionLabel
- \recoBkgdConstVec_{\signalRegionLabel}^\mathrm{CV} \big) \,,
\end{equation}
where, using similar reasoning and notation as in
Sec.~\ref{sec:sideband_data_stat_unc}, I have defined the covariance matrix
\begin{equation}
\label{eq:const_tot_with_data_unc}
\auxCovMat_{ \recoBkgdConstWithUncVec_\signalRegionLabel
  \recoBkgdConstWithUncVec_\signalRegionLabel }
= \auxCovMat_{ \recoBkgdConstVec_\signalRegionLabel
  \recoBkgdConstVec_\signalRegionLabel }
+ \auxCovMat_{ \dataRecoVec_\signalRegionLabel
  \dataRecoVec_\signalRegionLabel }
\end{equation}
to include the data statistical uncertainty $\auxCovMat_{
\dataRecoVec_\signalRegionLabel \dataRecoVec_\signalRegionLabel }$ on
$\dataRecoVec_\signalRegionLabel$.
A smaller value for $\chi^2_\mathrm{post}$ than for the
pre-constraint chi-squared metric
\begin{equation}
\chi^2_\mathrm{pre} = \big( \dataRecoVec_\signalRegionLabel
- \recoVec_{\signalRegionLabel}^\mathrm{CV} \big)^T \cdot
\auxCovMat_{ \recoWithUncVec_{\signalRegionLabel}
  \recoWithUncVec_{\signalRegionLabel} }^{-1}
\cdot \big( \dataRecoVec_\signalRegionLabel
- \recoVec_{\signalRegionLabel}^\mathrm{CV} \big) \,,
\end{equation}
where
\begin{equation}
\auxCovMat_{ \recoWithUncVec_\signalRegionLabel
  \recoWithUncVec_\signalRegionLabel }
= \auxCovMat_{ \recoVec_\signalRegionLabel
  \recoVec_\signalRegionLabel }
+ \auxCovMat_{ \dataRecoVec_\signalRegionLabel
  \dataRecoVec_\signalRegionLabel } \,,
\end{equation}
indicates a net improvement in agreement between the signal-region data and the
MC prediction following the constraint. A substantially larger
$\chi^2_\mathrm{post}$ value, on the other hand, may be symptomatic of
deficiencies in the simulation and/or poorly-designed sidebands, either of
which could necessitate adjustments to the analysis.

\subsection{Use in cross-section extraction}
\label{sec:use_in_xsec}

The constrained background prediction can be removed from the signal-region
data in each reconstructed bin to yield a vector of background-subtracted event
counts
\begin{equation}
\label{eq:bkgd_removal_ccbc}
\mathbf{\BkgdSubtractedRecoEvents}
= \dataRecoVec_\signalRegionLabel
- \bkgdVec^{\randVecCondLabel,\mathrm{CV}}_\signalRegionLabel
\end{equation}
needed as input for cross-section extraction. The other required inputs for
MicroBooNE's approach are the constrained total covariance matrix $\auxCovMat_{
\recoBkgdConstWithUncVec_\signalRegionLabel
\recoBkgdConstWithUncVec_\signalRegionLabel }$ from
Eq.~\ref{eq:const_tot_with_data_unc} (the data statistical uncertainty should
be included in this context), the detector response matrix ($\ResponseMatrix$,
see Eqs.~\ref{eq:migmat}~and~\ref{eq:response_matrix}) and the expected signal
event counts in each true bin
($\PredictedSignalEvtCount_\trueBinIdx^\mathrm{CV}$). The last two of these
require no changes after the background-only constraint since the central-value
signal prediction $\sigVec_\signalRegionLabel^\mathrm{CV}$ remains unchanged
from the nominal simulation.

Since the T2K extraction procedure allows for control sample bins to be
included in the likelihood fit, a background constraint that shares some
similarities to the CCBC is already included in their scheme automatically. The
formalism presented here offers no special advantage over the existing
technique.

For MINERvA-style analyses, since the systematic universes are evaluated on the
cross sections obtained during repeated unfoldings rather than on reconstructed
event counts (see Sec.~\ref{sec:multi_universe_extract}), the covariance
matrices needed to apply the CCBC may not be readily available.
If one assumes that they can be calculated prior to unfolding, then a
background constraint based upon them can be applied in the following way. When
re-extracting the cross section in each systematic universe, subtract a new
constrained background prediction
computed by making the substitutions
$\bkgdVec_{\signalRegionLabel}^{\randVecCondLabel,\mathrm{CV}} \to
\bkgdVec_{\signalRegionLabel}^{\randVecCondLabel,\UnivIdx}$,
$\bkgdVec_{\signalRegionLabel}^\mathrm{CV} \to
\bkgdVec_{\signalRegionLabel}^\UnivIdx$,
and
$\recoVec_{\controlSampleLabel}^\mathrm{CV} \to
\recoVec_{\controlSampleLabel}^\UnivIdx$ in Eq.~\ref{eq:ccbc_bkgd} and
Eq.~\ref{eq:bkgd_removal_ccbc}. Here the superscript $\UnivIdx$ indicates that
the relevant quantity is evaluated in the systematic universe of interest
$\UnivIdx$. Apart from this adjustment to background subtraction, the remainder
of the extraction workflow proceeds normally. The CCBC in this case serves as
an alternative to the various methods for fitting the background scale factors
$\bkgdScaleFactor_{\bkgdIdx}$ (see Sec.~\ref{sec:sidebands_exper}) used in many
MINERvA measurements.

\section{Summary}

As neutrino experimentalists continue to seek answers to compelling open
questions via increasingly precise measurements, corresponding improvements to
the precision of neutrino interaction models and simulations will be
essential. For accelerator-based experiments operating at \si{GeV} energies, a
sizeable literature of neutrino cross-section measurements already exists to
guide these improvements, and multiple experiments continue to expand the
available data sets at an increasing pace.

Cross-section extraction, a concept central to the large majority of neutrino
scattering measurements, describes a family of analysis techniques by which
event counts measured by a detector are converted into cross-section results
directly comparable to theoretical calculations. While the major approaches to
cross-section extraction, as typified herein by the standard strategies from
MINERvA, MicroBooNE, and T2K, are superficially similar and share a common
goal, their mathematical implementations contain significant differences in
unfolding method, uncertainty quantification, and interpretation of the
results in light of how flux shape modeling uncertainties are
handled~\cite{KochDolan}. In this paper, I have reviewed these similarities
and differences in the hope of spurring further community discussion on best
practices for the field.

I have also proposed two techniques intended to advance the state of the art
in cross-section extraction methods and thus enhance the power of future
measurements to inform neutrino interaction model development. The first
technique, blockwise unfolding, seeks to remedy the long-standing omission of
correlated uncertainties between separate kinematic distributions reported in
the same cross-section data release. The lack of information about such
correlations limits the full model discrimination potential of current
neutrino data sets, but an uncertainty treatment that makes them available is
readily achievable if planned ahead of time. Blockwise unfolding defines a way
of organizing and presenting neutrino cross-section measurements that allows
inter-distribution covariances to be reported, including the possibility of
doing so for distributions obtained from distinct analyses from the same
experiment.

The second technique proposed in this paper, the conditional covariance
background constraint (CCBC), seeks to improve the precision of the removal of
background events in the context of MicroBooNE-like cross-section analyses. In
many analyses performed by other experiments, background estimation based upon
a nominal simulation prediction is refined in a data-driven way using ancillary
measurements known as sidebands. While highly useful for improving the quality
of measurements in which background contamination is significant, this general
approach has not yet been adopted for any of MicroBooNE's differential cross
section results released to date. Recognizing that this may be due in part to a
lack of a clear way to incorporate the use of a sideband-based model constraint
in MicroBooNE's preferred method of cross-section extraction, I build upon an
approach first described in Ref.~\cite{uBEta} to construct the CCBC as a fully
compatible recipe. Although devised primarily to solve a problem specific to
MicroBooNE-style measurements, analyses adopting MINERvA-like cross-section
extraction may also potentially employ a version of the CCBC as an alternative
to the standard method based on fits of normalization scale factors.

\section{Acknowledgments}

I thank the organizers of the October 2023 NuXTract workshop at
CERN~\cite{nuxtract}, which helped to catalyze the writing of this article and
served as an excellent forum to explore related ideas. I also thank the members
of the MicroBooNE cross-section working group for helpful discussions.

This manuscript has been authored by Fermi Research Alliance, LLC under
Contract No. DE-AC02-07CH11359 with the U.S. Department of Energy, Office of
Science, Office of High Energy Physics.

\appendix
\section{Wiener-SVD regularization matrix}
\label{sec:Ac_wsvd}

In the Wiener-SVD method, the regularization matrix is calculated according to
the expression~\cite{WienerSVD}
\begin{equation}
\label{eq:wsvd_add_smear}
\AddSmearMatrix = C^{-1} \cdot \mathcal{V}_C \cdot W_C
\cdot \mathcal{V}_C^{T} \cdot C \,.
\end{equation}

The assisting matrix $C$ determines how $\AddSmearMatrix$ relates to a prior
prediction of the expected unfolded event counts. Common choices for $C$
include the identity matrix and forms based on estimation of derivatives using
finite differences. The matrix $\mathcal{V}_C$ is obtained in a two-step
procedure. First, Cholesky decomposition is applied to the inverse of a
covariance matrix
\begin{equation}
\CovMat^\mathrm{reco}_{\recoBinIdx \secondRecoBinIdx}
\equiv \mathrm{Cov}(\AllRecoEvents_\recoBinIdx,
\AllRecoEvents_\secondRecoBinIdx)
\end{equation}
representing the total statistical and systematic uncertainty on the measured
event counts. This yields
\begin{equation}
(\CovMat^\mathrm{reco})^{-1} = Q^T \cdot Q
\end{equation}
where $Q$ is a lower triangular matrix. Second, singular value decomposition is
applied to the matrix
\begin{equation}
G \equiv Q \cdot \ResponseMatrix \cdot C^{-1}
\end{equation}
to obtain
\begin{equation}
G = \mathcal{U}_C \cdot \WSVDDiagMatrix \cdot \mathcal{V}_C^T \,.
\end{equation}
Here $\mathcal{U}_C$ ($\mathcal{V}_C$) is an orthogonal matrix with a number of
rows equal to the number of reconstructed (true) bins, while $\WSVDDiagMatrix$
is a diagonal matrix.

To calculate the Wiener filter matrix $W_C$, a linear transformation
$\mathbf{g}$ of a vector $\pmb{\PredictedSignalEvtCount}$ of predicted signal
event counts in each true bin
\begin{equation}
\mathbf{g} \equiv \mathcal{V}_C^T \cdot C \cdot \pmb{\PredictedSignalEvtCount}
\end{equation}
is first evaluated. Define the symbol
\begin{equation}
h_\trueBinIdx \equiv g_\trueBinIdx \, \mathcal{D}_{\trueBinIdx \trueBinIdx} \,,
\end{equation}
where $g_\trueBinIdx$ is the element of $\mathbf{g}$ corresponding to the
\mbox{$\trueBinIdx$-th} true bin, and $\mathcal{D}_{\trueBinIdx \trueBinIdx}$
is the $\trueBinIdx$-th diagonal element of $\WSVDDiagMatrix$. Then one may
write the elements $w_{\trueBinIdx \secondTrueBinIdx}$ of $W_C$ in the form
\begin{equation}
w_{\trueBinIdx \secondTrueBinIdx}
= \frac{ h_\trueBinIdx^2 }{ h_\trueBinIdx^2 + 1 }
\, \delta_{\trueBinIdx \secondTrueBinIdx} \,,
\end{equation}
where $\delta_{\trueBinIdx \secondTrueBinIdx}$ is the Kronecker delta.

\bibliography{prd.bib}

\end{document}